\documentclass[a4paper,11pt]{article}
\usepackage{jcappub}

\usepackage{bm}
\usepackage{booktabs}
\usepackage{multirow}
\usepackage{listings}
\usepackage{xcolor}
\newcommand{\hicosmo}{\textsc{HIcosmo}}
\newcommand{\jax}{\textsc{JAX}}
\newcommand{\numpyro}{\textsc{NumPyro}}
\newcommand{\lcdm}{$\Lambda$CDM}
\newcommand{\wcdm}{$w$CDM}
\newcommand{\Omegam}{\Omega_{\rm m}}
\newcommand{\Omegab}{\Omega_{\rm b}}
\newcommand{\Omegar}{\Omega_{\rm r}}
\newcommand{\OmegaL}{\Omega_\Lambda}
\newcommand{\Omegak}{\Omega_k}

\begin{document}

\title{\hicosmo: a differentiable \jax-based framework for cosmology inference}

\author[a]{Jing-Zhao Qi,}
\author[a]{Jing-Fei Zhang}
\author[a,b,c,d,\ast]{and Xin Zhang\note[$\ast$]{Corresponding author.}}

\affiliation[a]{Department of Physics, College of Sciences, Northeastern University, Shenyang 110819, China}
\affiliation[b]{Liaoning Key Laboratory of Cosmology and Astrophysics, Northeastern University, Shenyang 110819, China}
\affiliation[c]{MOE Key Laboratory of Data Analytics and Optimization for Smart Industry, Northeastern University, Shenyang 110819, China}
\affiliation[d]{National Frontiers Science Center for Industrial Intelligence and Systems Optimization, Northeastern University, Shenyang 110819, China}

\emailAdd{qijingzhao@neu.edu.cn, jfzhang@neu.edu.cn, zhangxin@neu.edu.cn}

\abstract{
The Stage IV cosmological surveys, such as Euclid, LSST, DESI, and SKA, will deliver observational data of unprecedented volume, calling for efficient and reliable inference tools.
This paper presents \hicosmo\ (High-performance Inference for Cosmology), an open-source \jax-based framework for cosmology inference.
In \hicosmo, the forward model, distance integrals, likelihood evaluations, posterior sampling, and Fisher forecasts are all built from \jax\ primitives, so that gradients and Hessians of the log-likelihood are obtained directly by automatic differentiation, without any finite-difference approximation.
The framework implements the \lcdm, \wcdm, $w_0 w_a$CDM, and interacting dark-energy models, and provides likelihoods for Type Ia supernovae (Pantheon+, DES-SN5YR, Union3), baryon acoustic oscillations (DESI~DR1/DR2, SDSS), Planck~2018 distance priors, local $H_0$ measurements, and strong-lensing time delays.
Its scope is restricted to background cosmology, with Boltzmann solvers and full perturbation-level likelihoods left to external tools.
We validate \hicosmo\ against the reference implementation of each likelihood and against Cobaya. $\chi^2$ values agree to absolute differences of $10^{-6}$--$10^{-2}$, and the marginalized constraints from the two codes differ by less than $0.2\sigma$ in every analysis tested.
Leveraging just-in-time compilation and automatic differentiation, \hicosmo\ achieves about $8.7\times$ the end-to-end sampling throughput of Cobaya on CPU. As the dataset grows to survey scale, GPU acceleration over CPU reaches up to $20\times$.
As applications, we present multi-probe \lcdm\ joint constraints, dark-energy equation-of-state constraints, and Fisher forecasts for six 21\,cm intensity-mapping surveys, including SKA1, MeerKAT, BINGO, Tianlai, and CHIME.
}

\maketitle

\section{Introduction}
\label{sec:introduction}

Nearly a century of cosmological observations has established the six-parameter $\Lambda$ cold dark matter (\lcdm) model as the standard model of cosmology, supported by a wide range of data \cite{SupernovaSearchTeam:1998fmf,SupernovaCosmologyProject:1998vns,WMAP:2003elm,SDSS:2003eyi,SDSS:2004wzw}, with the \textit{Planck}\ satellite \cite{Planck:2015fie,Planck2018} testing the model to unprecedented precision.
As the measurements have been improved, however, inconsistencies have emerged between several key parameters inferred from the early and the late Universe \cite{Planck2018,Riess:2019cxk,DiValentino:2021izs,Guo:2018ans}.
The most prominent is the Hubble constant tension, the significant difference between the $H_0$ value inferred from \textit{Planck}\ data and that inferred from local SNe~Ia calibrated with the Cepheid distance ladder.
The $S_8$ tension between cosmic-shear surveys and \textit{Planck}~2018 \cite{Heymans:2020gsg}, as well as the enhanced lensing amplitude in the CMB power spectrum \cite{DiValentino:2019qzk}, also challenges the standard model.
These tensions suggest that our understanding of the Universe within the standard model may be incomplete, and their clarification requires further investigation.

To clarify these tensions and further improve the parameter measurements, a new generation of cosmological surveys is under construction or being planned.
Euclid \cite{Euclid} will measure the shapes and redshifts of more than a billion galaxies, and LSST \cite{LSST} is expected to detect $\mathcal{O}(10^5)$ Type Ia supernovae over its ten-year survey.
The BAO analysis of DESI \cite{DESI2024} has already compressed the spectra of six million tracers into 12 measurements of $D_M/r_d$ and $D_H/r_d$ at seven effective redshifts.
21\,cm intensity-mapping (IM) experiments will further provide three-dimensional maps of large-scale structure in the radio band, with the SKA1 Band~1 survey alone covering $20{,}000$~deg$^2$ over $0.3<z<3$.
Data on this scale place new demands on the efficiency and reliability of cosmological inference tools.

The computational cost of cosmological inference depends largely on the level to which the data are compressed.
At the background level, likelihoods operate on distances, distance ratios, and distance moduli, and the parameter space for \lcdm\ or $w_0 w_a$CDM typically has only $d\sim 5$--$7$ dimensions.
Probes at this level have been used extensively to constrain cosmological parameters and to test the theory of gravity and fundamental principles of cosmology, with data including CMB, BAO, SNe~Ia, strong-lensing time delays, gravitational wave standard sirens, fast radio bursts, and other cosmological probes \cite{Cao2017AA,Qi2021JCAP,Qi2021MNRAS,Qi2023PRD,Qi2025ApJ,Wang:2018lun,Zhang:2018byx,Zhang:2019ylr,Zhang:2019loq,Wang:2019tto,Zhao:2019gyk,Wang:2021srv,Zhao:2020ole,Qiu:2021cww,Zhao:2022bpd,Wang:2021kxc,Jin:2022qnj,Jin:2023sfc,Jin:2021pcv,Song:2022siz,Zhao:2022yiv,Song:2025bio,Zhang:2023gye,Zhang:2024rra}, and mature codes such as Cobaya \cite{Cobaya2021} handle such analyses comfortably on an ordinary workstation.
The situation at the perturbation level is very different.
Weak-lensing and galaxy-clustering $3\times 2$pt analyses require repeated evaluations of the matter power spectrum $P(k,z)$ by Boltzmann solvers such as CAMB \cite{CAMB} or CLASS \cite{CLASS}, the parameter dimension reaches $d\sim 30$--$50$ for a single survey, and multi-survey combinations easily exceed 150 parameters \cite{Piras2024}.
At such dimensionality, the efficiency of random-walk Metropolis--Hastings (MH) sampling degrades rapidly \cite{Betancourt2017}.
Piras et al. \cite{Piras2024} estimate that a traditional nested-sampling analysis of a 157-parameter Stage~IV problem would cost about $10^4$ CPU core-days.

Gradient-based samplers offer a way out of this high-dimensional sampling problem.
The No-U-Turn Sampler (NUTS) \cite{Hoffman2014} is an adaptive variant of Hamiltonian Monte Carlo (HMC), whose mixing time for smooth high-dimensional targets scales roughly as $d^{1/4}$, far better than the $d$ scaling of random-walk MH \cite{Betancourt2017}.
The difficulty is that the mature cosmological pipelines, including CosmoMC \cite{Lewis2002}, Cobaya \cite{Cobaya2021}, emcee \cite{ForemanMackey2013}, and MontePython \cite{MontePython}, are not implemented as differentiable programs.
Gradients of the log-likelihood can be approximated by finite differences, but each gradient then costs $O(d)$ extra likelihood evaluations and is sensitive to the choice of step size.

The development of automatic differentiation (AD) \cite{Griewank2008} and \jax\ \cite{JAX2018} has changed this situation.
\jax\ brings automatic differentiation, just-in-time (JIT) compilation, vectorization, and a unified CPU/GPU/TPU execution interface to the Python ecosystem on which most cosmological codes rely.
Piras et al. \cite{Piras2024} have demonstrated the potential of this approach at the perturbation level, where their \jax-based Stage~IV $3\times 2$pt pipeline completed a 157-parameter inference in about 8 days on 24 GPUs, an estimated $545\times$ speedup over the equivalent nested-sampling analysis.
Several \jax-native cosmological tools have appeared.
jax-cosmo \cite{JAXCosmo} implements differentiable distance calculations and weak-lensing power spectra, while CosmoPower \cite{CosmoPower} replaces the Boltzmann solver with neural-network emulators of $C_\ell$ and $P(k)$, whose forward pass and gradients have been ported to \jax\ \cite{Piras2024}.
At the background level, however, there is still no systematically validated \jax-native inference pipeline that provides differentiable likelihoods, NUTS sampling, and AD-based Fisher forecasts within a single framework, with end-to-end comparisons against the reference codes commonly used in cosmology at every step.
This work fills that gap.

In this paper, we present \hicosmo\ (\textbf{H}igh-performance \textbf{I}nference for \textbf{Cosmo}logy), a differentiable \jax-based framework for background-cosmology inference.
\hicosmo\ implements four background models, namely \lcdm, \wcdm, $w_0 w_a$CDM, and an interacting dark-energy model.
The supported likelihoods cover Type Ia supernova samples (Pantheon+ \cite{Scolnic2022}, DES-SN5YR \cite{DESY5}, Union3 \cite{Rubin2023}), BAO measurements (DESI~DR1 \cite{DESI2024}, DESI~DR2 \cite{DESIDR2}, SDSS), Planck~2018 distance priors, the SH0ES local $H_0$ measurement, and strong-lensing time delays from H0LiCOW \cite{Wong2020}.
Because every component is written in \jax, the \numpyro\ NUTS sampler \cite{Phan2019} and AD-based Fisher forecasts connect directly to the same likelihood, without switching inference frameworks.
We validate \hicosmo\ systematically against the official codes and Cobaya, and find that the parameter constraints from the two codes agree within statistical uncertainties for the tested \lcdm, \wcdm, and $w_0 w_a$CDM configurations.

The paper is organized as follows.
Section~\ref{sec:architecture} describes the layered architecture of \hicosmo, its \jax\ implementation, and the supported observational probes.
Section~\ref{sec:validation} validates \hicosmo\ against reference codes and published constraints.
Section~\ref{sec:performance} presents performance benchmarks.
Section~\ref{sec:applications} demonstrates multi-probe \lcdm\ joint constraints, dark-energy equation-of-state constraints, and Fisher forecasts for 21\,cm intensity-mapping surveys.
Section~\ref{sec:discussion} places \hicosmo\ in the landscape of existing inference tools, discusses its limitations and the extension to perturbation-level inference, and concludes.

\section{Architecture and implementation}
\label{sec:architecture}

\begin{figure}[htbp]
    \centering
    \includegraphics[width=0.95\textwidth]{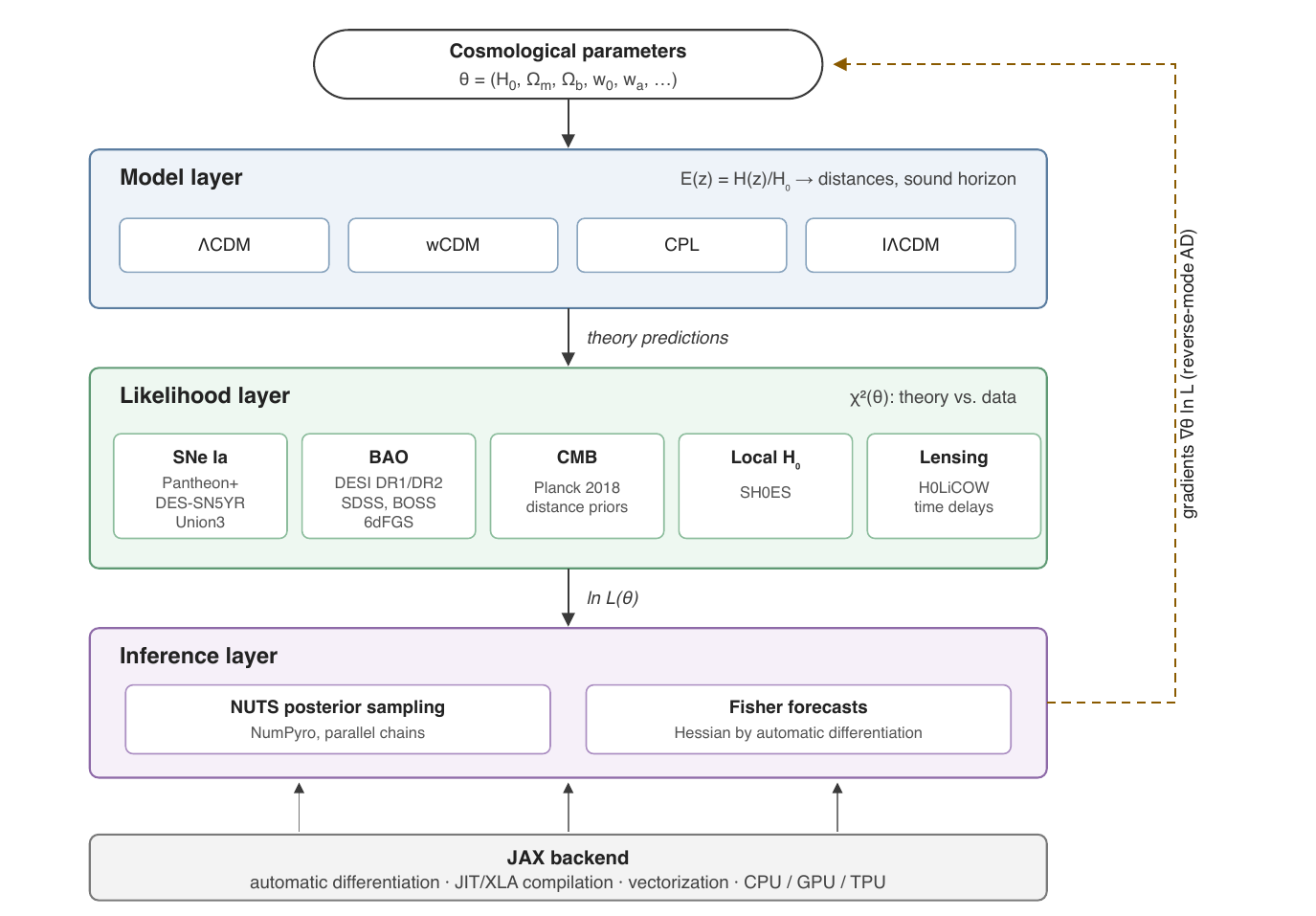}
    \caption{Layered architecture of \hicosmo.
    The model layer computes $E(z)$ and the derived distances and sound horizon from the cosmological parameters $\theta$ (\lcdm, \wcdm, $w_0 w_a$CDM, I\lcdm).
    The likelihood layer compares the theoretical predictions with the data of each observational probe and returns $\chi^2(\theta)$.
    The inference layer consumes the resulting log-likelihood for NUTS posterior sampling and for Fisher forecasts based on the AD Hessian.
    All computations are supported by the \jax\ backend (automatic differentiation, JIT/XLA compilation, vectorization, and unified CPU/GPU execution).
    Solid arrows indicate the forward data flow, and the dashed arrow indicates the gradients $\nabla_\theta \ln\mathcal{L}$ returned to the sampler by reverse-mode automatic differentiation.}
    \label{fig:architecture}
\end{figure}

This section describes the layered architecture of \hicosmo\ and its \jax\ implementation.
As shown in figure~\ref{fig:architecture}, the framework consists of a model layer, a likelihood layer, and an inference layer.
The model layer computes theoretical predictions from the cosmological parameters, the likelihood layer compares these predictions with observational data, and the inference layer receives the resulting log-likelihood and its derivatives to perform posterior sampling and Fisher forecasting.
Every operation from the cosmological parameters to the scalar log-likelihood is built from \jax\ primitives, so automatic differentiation, JIT compilation, and unified CPU/GPU execution are available throughout the pipeline.

\subsection{Design principles}
\label{sec:design}

The central design principle of \hicosmo\ is end-to-end differentiability.
As long as every step from the cosmological parameters to the log-likelihood can be traced by the \jax\ automatic-differentiation system, the gradient of the likelihood follows directly from applying the chain rule to the computation graph, with no step size to choose and no $O(d)$ overhead of coordinate-wise finite differences.
The cost of one gradient evaluation depends only on the size of the computation graph and amounts to about 2--5 likelihood evaluations \cite{Griewank2008}.
The NUTS sampler explores the posterior using gradients, Fisher forecasts obtain the Hessian by nested automatic differentiation, and adding a new model or likelihood requires no hand-written derivative code.

Achieving end-to-end differentiability requires a systematic rewrite of the numerical routines on which mature cosmological codes rely, because adaptive integrators, data-dependent control flow, and calls into external compiled libraries cannot be traced by \jax.
In \hicosmo, adaptive integrators are replaced by fixed-grid trapezoidal quadrature, loops are replaced by vectorized operations, and parameter-independent quantities such as inverse covariance matrices are precomputed once.
These changes affect only the numerical methods. The physical definitions of all likelihoods follow the literature.
Precision checks for each substitution are given alongside the corresponding component in this section, and end-to-end validation is presented in section~\ref{sec:validation}.

In addition to differentiability, the execution speed of the likelihood matters for MCMC.
The sampler calls the likelihood $10^5$--$10^6$ times, and the per-call overhead of the Python interpreter becomes significant at this volume.
The JIT compiler in \jax\ compiles the likelihood into optimized machine code on the first call, and every subsequent call executes the compiled result directly, bypassing the interpreter, and the one-time compilation cost is negligible over a full run.
The same code runs unchanged on CPUs or GPUs.

\subsection{Core foundation layer}

The core foundation layer provides the distance and integration calculations shared by all cosmological models.
Each model only needs to specify its dimensionless Hubble parameter $E(z) = H(z)/H_0$.
From this single function, the base class computes $H(z)$, the comoving distance, the transverse comoving distance, the luminosity distance $d_L(z)$, the angular-diameter distance $d_A(z)$, the distance modulus $\mu(z)$, and the sound horizon $r_s(z)$.
Constructing a new cosmological model therefore involves only the implementation of a new $E(z)$, and the downstream distance calculations do not need to be rebuilt.

The central quantity is the comoving-distance integral,
\begin{equation}
    D_C(z) = \frac{c}{H_0} \int_0^z \frac{dz'}{E(z')}.
    \label{eq:comoving_distance}
\end{equation}
Traditional implementations evaluate this integral with adaptive quadrature, but the control flow of an adaptive algorithm depends on run-time error estimates, which cannot be handled by the tracing compilation of \jax, and the routine is also a black box that exposes no derivative information to automatic differentiation.
\hicosmo\ instead uses fixed-step trapezoidal integration on a dense redshift grid with spacing $\Delta z \approx 5\times 10^{-4}$ (2,000--16,000 points depending on the redshift range of the probe).
Table~\ref{tab:integration_precision} compares this cumulative-trapezoid path, which is the one actually executed inside the MCMC likelihoods, with \texttt{scipy} adaptive quadrature.
Over the cosmologically relevant redshift range the relative deviation stays below $2\times 10^{-7}$, far smaller than the typical posterior uncertainties of $10^{-2}$--$10^{-1}$, so the integration error is negligible in all analyses of this paper.

\begin{table}[htbp]
\centering
\caption{Precision of the numerical integration: relative deviation of the comoving distance between the fixed-step cumulative-trapezoid integration of \hicosmo\ (8192-point grid spanning $0\le z\le 5$, the configuration used by the BAO likelihoods) and \texttt{scipy} adaptive quadrature.
Test cosmology: $H_0=67.36$~km/s/Mpc, $\Omegam=0.315$, $\Omegak=0$, with $\Omegar$ from eq.~(\ref{eq:omega_r}).}
\label{tab:integration_precision}
\begin{tabular}{cccc}
\toprule
$z$ & scipy [Mpc] & \hicosmo\ [Mpc] & Relative deviation \\
\midrule
0.1 & 434.36 & 434.36 & $1.5\times10^{-7}$ \\
0.5 & 1952.45 & 1952.45 & $1.7\times10^{-8}$ \\
1.0 & 3402.96 & 3402.96 & $7.0\times10^{-9}$ \\
2.0 & 5314.32 & 5314.32 & $2.6\times10^{-9}$ \\
5.0 & 7948.24 & 7948.24 & $7.7\times10^{-9}$ \\
\bottomrule
\end{tabular}
\end{table}

The transverse comoving distance $D_M(z)$ depends on the spatial curvature:
\begin{equation}
    D_M(z) = \begin{cases}
        \frac{D_H}{\sqrt{\Omegak}} \sinh\left(\frac{\sqrt{\Omegak} D_C}{D_H}\right) & \Omegak > 0 \\
        D_C & \Omegak = 0 \\
        \frac{D_H}{\sqrt{-\Omegak}} \sin\left(\frac{\sqrt{-\Omegak} D_C}{D_H}\right) & \Omegak < 0
    \end{cases}\,,
    \label{eq:transverse_distance}
\end{equation}
where $D_H = c/H_0$ is the Hubble distance.
The distance calculations in \hicosmo\ support non-flat geometries with $\Omegak$ as a free parameter.

For CMB-related analyses, the radiation density parameter $\Omegar$ must be treated consistently with $H_0$ and the effective number of neutrino species $N_{\rm eff}$.
Otherwise, small inconsistencies in $\Omegar$ propagate through $D_M(z_*)$ into the Planck distance priors and produce systematic biases.
In \hicosmo, $\Omegar$ is derived strictly from the CMB temperature $T_{\rm cmb}$, $H_0$, and $N_{\rm eff}$,
\begin{equation}
    \Omegar = \frac{2.4728 \times 10^{-5}}{h^2}
    \left(\frac{T_{\rm cmb}}{2.7255\,\rm K}\right)^4
    (1 + 0.22711 N_{\rm eff}),
    \label{eq:omega_r}
\end{equation}
where $h = H_0/(100\,\rm km/s/Mpc)$.
All cosmological models in the framework share this expression, guaranteeing a single consistent definition of $\Omegar$ throughout all analyses.
For the same reason, the physical constants ($c$, $G$), the CMB reference values ($T_{\rm cmb}=2.7255$~K, $N_{\rm eff}=3.046$), and the Planck 2018 best-fit default parameters are maintained centrally in a single constants module.

\subsection{Cosmological model layer}

The cosmological model layer provides the concrete implementations of the background models.
Each model inherits from the base class and only specifies $E(z)$, after which all distance calculations become available automatically.
In this paper we focus on the following four widely studied background models.

\paragraph{\lcdm\ model.} The dimensionless Hubble parameter is
\begin{equation}
    E^2(z) = \Omegam(1+z)^3 + \Omegar(1+z)^4 + \Omegak(1+z)^2 + \OmegaL,
    \label{eq:E_z_lcdm}
\end{equation}
where $\OmegaL = 1 - \Omegam - \Omegar - \Omegak$ is fixed by the closure relation.

The \lcdm\ implementation also provides several derived quantities needed in the analyses.
The sound horizon at the baryon drag epoch $r_d\equiv r_s(z_d)$ is computed from the Eisenstein--Hu fitting formula \cite{Eisenstein1998}. The raw formula deviates from Boltzmann solvers \cite{CAMB, CLASS} by $1$--$2\%$; following the DESI analysis \cite{DESI2024}, \hicosmo\ calibrates $r_d$ against CAMB, reducing the residual to about $0.01\%$. CAMB can also be called directly when an exact Boltzmann $r_d$ is required (section~\ref{sec:cobaya_validation}). $r_d$ and the last-scattering sound horizon $r_s(z_*)$ used by the CMB likelihood differ by about $1\%$ and are kept as independent quantities in the code. The derived quantities commonly used in BAO analyses, $r_d$ and $H_0 r_d$, are computed automatically by the framework.

\paragraph{\wcdm\ model.} Starting from \lcdm, the dark-energy equation of state $w$ is released and allowed to deviate from $-1$,
\begin{align}
    E^2(z) &= \Omegam(1+z)^3 + \Omegar(1+z)^4 + \Omegak(1+z)^2 \notag \\
    &\quad + \Omega_{\rm DE}(1+z)^{3(1+w)}.
    \label{eq:E_z_wcdm}
\end{align}
For $w = -1$ the model reduces to \lcdm.

\paragraph{$w_0 w_a$CDM model.} The Chevallier--Polarski--Linder (CPL) parametrization \cite{Chevallier2001,Linder2003} allows the dark-energy equation of state to evolve with redshift,
\begin{equation}
    w(z) = w_0 + w_a \frac{z}{1+z},
    \label{eq:cpl_w}
\end{equation}
with the corresponding expansion rate
\begin{align}
    E^2(z) &= \Omegam(1+z)^3 + \Omegar(1+z)^4 + \Omegak(1+z)^2 \notag \\
    &\quad + \Omega_{\rm DE}(1+z)^{3(1+w_0+w_a)} \exp\left[-\frac{3w_a z}{1+z}\right].
    \label{eq:E_z_cpl}
\end{align}

\paragraph{I\lcdm\ model.} The interacting vacuum energy model considers energy exchange between dark matter and dark energy \cite{Li:2011ga,Zhang:2012sya,Li:2015vla,Guo:2017deu,Li:2018ydj,Guo:2018gyo,Feng:2017usu,Fu:2011ab,Li:2024qso,Li:2025owk,Li:2026xaz}, with the continuity equations
\begin{equation}
    \dot{\rho}_c + 3H\rho_c = Q, \quad
    \dot{\rho}_\Lambda + 3H(1+w)\rho_\Lambda = -Q,
    \label{eq:ilcdm}
\end{equation}
where the interaction term is taken as $Q = 3\beta H \rho_c$ and $\beta$ is a dimensionless coupling parameter.
For $w=-1$ the background admits the closed-form solution
\begin{equation}
    E^2(z) = \OmegaL + \frac{\beta\,\Omega_c}{\beta-1} + \frac{\Omega_c}{1-\beta}(1+z)^{3(1-\beta)} + \Omegab(1+z)^3 + \Omegar(1+z)^4,
    \label{eq:ilcdm_Ez}
\end{equation}
where $\Omega_c = \Omegam - \Omegab$ is the cold-dark-matter density. The constant term absorbs the energy transferred into the dark-energy component, and the expression reduces to \lcdm\ as $\beta \to 0$.
I\lcdm\ also serves as a test of the framework's extensibility. Although the model introduces energy exchange between dark matter and dark energy and its background evolution departs from the standard form, all computations are realized by simply adding the corresponding $E(z)$.

\subsection{Likelihood layer}

The likelihood layer computes the $\chi^2$ between the model predictions and the observational data.
Each likelihood module performs only this comparison, and all theoretical predictions come from the model layer.
Because the likelihood modules contain no cosmological calculations, any discrepancy between \hicosmo\ and a reference implementation can only arise from the $\chi^2$ construction or the input data.

\paragraph{Pantheon+ Type Ia supernovae.} The Pantheon+ sample \cite{Scolnic2022} contains 1701 Type Ia supernovae and is the largest standardized SN~Ia compilation to date.
The likelihood is built from the distance-modulus residuals,
\begin{equation}
    \chi^2 = \Delta\boldsymbol{\mu}^T \mathbf{C}^{-1} \Delta\boldsymbol{\mu},
\end{equation}
where $\Delta\boldsymbol{\mu} = \boldsymbol{\mu}_{\rm obs} - \boldsymbol{\mu}_{\rm th}$ and $\mathbf{C}$ is the full covariance matrix including statistical and systematic uncertainties ($1701\times 1701$ as released, the standard $z>0.01$ cut used throughout this paper retains 1580 SNe and the corresponding $1580\times 1580$ block).
The absolute magnitude $M_B$ is marginalized analytically, so that the $H_0$ information from other probes in a joint analysis is not affected by an $M_B$ prior.
For analyses anchored to the local distance ladder, the SH0ES Cepheid calibration can be switched on independently as an option.

\paragraph{DESI 2024 BAO.} The first DESI data release \cite{DESI2024} provides 12 BAO measurements in seven redshift bins ($0.295 \le z_{\rm eff} \le 2.330$).
Five bins constrain $D_M(z)/r_d$ and $D_H(z)/r_d$ separately, where $D_H(z) = c/H(z)$ is the Hubble distance, while the BGS ($z_{\rm eff}=0.295$) and QSO ($z_{\rm eff}=1.491$) bins constrain the volume-averaged distance $D_V(z)/r_d$, with $D_V = [z D_M^2 D_H]^{1/3}$.
The likelihood reads
\begin{equation}
    \chi^2 = \Delta\mathbf{d}^T \, \mathbf{C}^{-1} \, \Delta\mathbf{d},
\end{equation}
where $\Delta\mathbf{d}$ is the residual vector of all 12 measurements and $\mathbf{C}$ is the corresponding covariance matrix.
\hicosmo\ uses the original covariance matrices released by the DESI collaboration, which include intra- and inter-bin correlations.

\paragraph{Planck 2018 distance priors.} A full CMB likelihood requires a Boltzmann solver to compute the CMB power spectra, which is computationally expensive and outside the scope of \hicosmo.
For analyses concerned only with background parameters, the CMB information can be compressed into three derived parameters \cite{Chen2019},
\begin{align}
    R &= \sqrt{\Omegam H_0^2} \frac{D_M(z_*)}{c}, \\
    l_a &= \pi \frac{D_M(z_*)}{r_s(z_*)}, \\
    \omega_b &= \Omegab h^2,
\end{align}
where $z_* \approx 1089.9$ is the redshift of last scattering.
According to ref.~\cite{Chen2019}, this three-parameter compression retains more than 95\% of the background information of the full Planck likelihood on $H_0$ and $\Omegam$, while removing the need for a Boltzmann solver and greatly reducing the evaluation cost.
\hicosmo\ adopts the Planck 2018 TT,TE,EE+lowE distance priors of ref.~\cite{Chen2019}, with mean vector
\begin{equation}
    (R,\; l_a,\; \omega_b) = (1.750235,\; 301.4707,\; 0.02235976)\,,
\end{equation}
and inverse covariance matrix
\begin{equation}
    \mathbf{C}^{-1} =
    \begin{pmatrix}
        9.439240\times 10^{4} & -1.360491\times 10^{3} & 1.664517\times 10^{6} \\
        -1.360491\times 10^{3} & 1.614349\times 10^{2} & 3.671618\times 10^{3} \\
        1.664517\times 10^{6} & 3.671618\times 10^{3} & 7.971918\times 10^{7}
    \end{pmatrix},
    \label{eq:planck_invcov}
\end{equation}
which together fully specify the likelihood.
Evaluating $R$ and $l_a$ requires $z_*$ and $r_s(z_*)$.
\hicosmo\ computes $z_*$ from the Hu--Sugiyama fitting formula \cite{HuSugiyama1996} and $r_s(z_*)$ from the Eisenstein--Hu sound-speed integral \cite{Eisenstein1998}, and calibrates both against CAMB in the $(\Omegab h^2, \Omegam h^2)$ plane following the standard practice of the BAO literature \cite{Aubourg2015, DESI2024}, reaching a calibration precision better than $2\times 10^{-5}$.
After calibration, the residuals in $R$ and $l_a$ are below $0.004\,\sigma_R$ and $0.03\,\sigma_{l_a}$, respectively.

\paragraph{H0LiCOW strong-lensing time delays.} The H0LiCOW collaboration \cite{Wong2020} measured $H_0$ independently of the distance ladder by modeling the time delays of six strongly lensed systems.
The time-delay distance $D_{\Delta t}$ is related to angular-diameter distances by
\begin{equation}
    D_{\Delta t} = (1+z_d)\, \frac{D_d D_s}{D_{ds}},
\end{equation}
where $D_d$, $D_s$, and $D_{ds}$ are the angular-diameter distances to the lens, to the source, and between the lens and the source, respectively.

\paragraph{SH0ES distance ladder.} The SH0ES collaboration \cite{Riess2022} calibrates SNe~Ia with Cepheids and provides a high-precision local $H_0$ measurement.
In \hicosmo\ this is implemented as a Gaussian prior on $H_0$,
\begin{equation}
    \chi^2 = \frac{(H_0 - 73.04)^2}{1.04^2}.
\end{equation}

\paragraph{DES Year 5 Type Ia supernovae.} The DES five-year supernova sample \cite{DESY5} combines photometrically classified DES supernovae with external low-redshift calibrators.
The Dovekie cross-calibration release used here provides 1820 bias-corrected distance moduli covering $0.025 < z < 1.15$, standardized with BBC + SALT3.
The $\chi^2$ has the same structure as Pantheon+, with the absolute-magnitude offset $M$ again marginalized analytically.
The difference is that the data are released as bias-corrected distance moduli $\mu$ rather than corrected apparent magnitudes $m_{b,\text{corr}}$, and the inverse covariance matrix is provided directly in compressed form, so no matrix inversion is needed at run time.

\paragraph{Union3 Type Ia supernovae.} The Union3 compilation \cite{Rubin2023} contains 2087 Type Ia supernovae standardized with the UNITY1.5 Bayesian hierarchical framework rather than the traditional Tripp formula.
The data are released as a compressed Gaussian approximation with 22 binned distance moduli and a $22\times 22$ inverse covariance matrix, covering $0.05 < z < 2.26$.
Because UNITY handles the absolute magnitude internally, no $M_B$ parameter enters the cosmological fit.

\paragraph{DESI DR2 BAO.} The second DESI data release \cite{DESIDR2} extends the 12 data points of DR1 to 13 points in seven redshift bins.
In particular, the QSO bin ($z_{\rm eff}=1.484$) is split from a single $D_V/r_d$ measurement into separate $D_M/r_d$ and $D_H/r_d$ constraints.
The DR2 likelihood in \hicosmo\ shares the same implementation as DR1, differing only in the input data, and the various treatments of $\Omegab$ (fixed, BBN prior, or free) apply equally.

\paragraph{Likelihood combination.} Multiple likelihoods are combined by direct addition in log-likelihood space, so a joint analysis is the sum of the individual log-likelihood contributions and the differentiability of each term is preserved automatically.
Table~\ref{tab:datasets} lists all observational datasets currently supported by \hicosmo.

\begin{table}[htbp]
\centering
\caption{Observational datasets supported by \hicosmo.
All likelihoods are implemented in pure \jax\ and support JIT compilation.}
\label{tab:datasets}
\begin{tabular}{lllc}
\hline\hline
Probe & Dataset & Reference & $N_{\rm data}$ \\
\hline
SNe Ia & Pantheon+ & \cite{Scolnic2022} & 1701 \\
       & Pantheon+ + SH0ES & \cite{Scolnic2022,Riess2022} & 1701 \\
       & DES-SN5YR (Dovekie) & \cite{DESY5} & 1820 \\
       & Union3 & \cite{Rubin2023} & 22 bins \\
\hline
BAO & DESI 2024 (DR1) & \cite{DESI2024} & 12 \\
    & DESI 2025 (DR2) & \cite{DESIDR2} & 13 \\
    & SDSS DR12/DR16 & \cite{BOSS:2016wmc,eBOSS:2020yzd} & 6/8 \\
    & BOSS DR12, 6dFGS & \cite{BOSS:2016wmc,Beutler:2011hx} & 4/1 \\
\hline
CMB & Planck 2018 dist.\ priors & \cite{Chen2019} & 3 \\
\hline
$H_0$ & SH0ES & \cite{Riess2022} & 1 \\
\hline
Strong lensing & H0LiCOW & \cite{Wong2020} & 6 \\
\hline\hline
\end{tabular}
\end{table}

\subsection{Inference and forecasting}
\label{sec:inference}

The inference layer receives the scalar log-likelihood and its derivatives and provides two modes of operation. The first is posterior sampling with NUTS \cite{Hoffman2014} as implemented in \numpyro\ \cite{Phan2019}; the second is Fisher-matrix forecasting via the AD Hessian.

NUTS simulates Hamiltonian dynamics on the negative log-posterior surface, with the required gradients supplied directly by \jax\ rather than by finite-difference approximations.
Compared with random-walk Metropolis--Hastings, NUTS selects the trajectory length automatically through the U-turn criterion and uses the gradient information to reduce the autocorrelation time of the chains.
Parallel chains are distributed over CPU cores or GPUs by \jax.
All analyses in this paper use the same default NUTS configuration, 8 parallel chains, 20{,}000 total samples, 1{,}000 warmup steps in total distributed over the chains, a target acceptance rate of 0.8, and a maximum tree depth of 10.
During warmup the sampler adapts the step size and mass matrix to the local posterior geometry, and the configuration is held fixed during the production phase.
Convergence is assessed automatically through the Gelman--Rubin statistic $\hat{R}$ and the effective sample size (ESS), and we adopt $\hat{R}<1.01$ as the convergence criterion.

Fisher forecasting uses the same likelihood as the MCMC sampling, only with a different operation applied to it.
The Fisher matrix is defined as
\begin{equation}
    F_{ij} = -\left\langle \frac{\partial^2 \ln \mathcal{L}}{\partial \theta_i \partial \theta_j} \right\rangle,
\end{equation}
whose inverse gives the Cram\'er--Rao lower bound on the parameter covariance and is the standard tool for forecasting the constraining power of future surveys.
Conventional implementations compute the Fisher matrix with central finite differences, which requires $(n+1)^2$ likelihood evaluations per matrix and is highly sensitive to the step size.
In \hicosmo, the Fisher matrix is obtained by taking the AD Hessian of the log-likelihood directly, and the remaining error sources are floating-point round-off and the numerical approximations already present in the forward model.
A quantitative comparison of the two approaches is given in section~\ref{sec:performance}.

The Fisher module of \hicosmo\ also contains a power-spectrum forecasting pipeline for 21\,cm intensity-mapping (IM) surveys, with built-in configuration parameters for several planned radio facilities. The forecasts for SKA1, MeerKAT, BINGO, Tianlai, and CHIME are presented in section~\ref{sec:fisher_21cm}.
These forecasts demonstrate that the same differentiable likelihoods serving the existing SN/BAO/CMB analyses extend directly to forecasting for next-generation surveys.

The three-layer architecture exposes a concise Python interface to the user. The following code performs a complete posterior analysis of \lcdm\ with Pantheon+ supernovae.
\begin{lstlisting}
from hicosmo.models import LCDM
from hicosmo.likelihoods import PantheonPlus
from hicosmo.samplers import MCMC

likelihood = PantheonPlus(LCDM)
params = {'H0': (67, 60, 80), 'Omega_m': (0.3, 0.1, 0.5)}
mcmc = MCMC(params, likelihood, chain_name='lcdm_sn')
mcmc.run(num_samples=20000)
\end{lstlisting}
Four lines of code complete model instantiation, likelihood construction, parameter specification, and NUTS sampling. Switching cosmological models requires only replacing \texttt{LCDM} with \texttt{wCDM} or \texttt{CPL}, and adding BAO constraints requires only combining likelihood objects with \texttt{CombinedLikelihood}. The inference layer handles gradient computation, multi-chain parallelization, and convergence diagnostics automatically.

\section{Parameter constraints and cross-validation}
\label{sec:validation}

In this section we present the parameter constraints obtained with \hicosmo\ under \lcdm, \wcdm, and $w_0 w_a$CDM, and systematically cross-validate each result against Cobaya \cite{Cobaya2021}.
The validation first compares each individual likelihood point by point with its reference implementation, and then compares the full posterior distributions under matched data and parameter choices, testing the likelihood and the sampler together.

Unless stated otherwise, all MCMC analyses in this paper use the default configuration of section~\ref{sec:inference} (8 parallel chains with 20{,}000 total samples and 1{,}000 warmup steps in total), while the $w_0 w_a$CDM analyses use longer runs with 50{,}000 samples.
BAO analyses use the CAMB-calibrated Eisenstein--Hu sound horizon $r_d$ described in section~\ref{sec:architecture}.
In the Cobaya cross-validation of this section, Cobaya computes $r_d$ with the full CAMB Boltzmann solver, and the two $r_d$ values agree to about $0.01\%$, so the residual differences originate only from the samplers and the numerical implementations, making the comparison a controlled one.

The likelihood-level comparison is performed point by point on identical inputs.
The deviations of \hicosmo\ from the reference implementations are below $0.01$ in the Pantheon+ $\chi^2$, below $10^{-4}$ in the DESI BAO log-likelihood, and below $10^{-6}$ in the Planck distance-prior likelihood.
These deviations are far smaller than the $\Delta\chi^2 \sim 1$ corresponding to the $1\sigma$ posterior contour, so any difference that appears later at the posterior level should not be attributed to the likelihood implementations themselves.

Table~\ref{tab:single_probe} summarizes the individual probe constraints under the \lcdm\ assumption (68\% credible intervals), all obtained on real observational data with the default sampling configuration of section~\ref{sec:inference}.

\begin{table}[htbp]
\centering
\caption{Single-probe \lcdm\ parameter constraints (68\% credible intervals).
All analyses use \hicosmo\ with the \numpyro\ NUTS sampler, 8 parallel chains, and 20,000 total samples.
$H_0$ is in km/s/Mpc.
$^a$~Prior-dominated (flat posterior after analytic $M_B$ marginalization).
$^b$~$\Omegab h^2$ fixed to the Planck 2018 best-fit value $0.02237$ for the computation of $r_d$.}
\label{tab:single_probe}
\begin{tabular}{lcc}
\toprule
Probe & $H_0$ & $\Omegam$ \\
\midrule
Planck 2018 & $67.50 \pm 0.61$ & $0.317 \pm 0.008$ \\
DESI BAO$^b$ & $68.99 \pm 0.71$ & $0.295 \pm 0.015$ \\
Pantheon+ & ---$^a$ & $0.333 \pm 0.019$ \\
SH0ES + SNe & $73.05 \pm 1.04$ & $0.333 \pm 0.018$ \\
H0LiCOW & $73.33 \pm 1.61$ & $0.338 \pm 0.130$ \\
\bottomrule
\end{tabular}
\end{table}

Table~\ref{tab:single_probe} displays the complementarity of the probes clearly.
The Planck CMB distance priors give the tightest single-probe constraints ($0.9\%$ on $H_0$ and $2.5\%$ on $\Omegam$), because the three derived quantities $(l_A, R, z_*)$ jointly pin down the high-redshift expansion geometry.
With $\Omegab h^2$ fixed to the Planck best fit, DESI BAO translates the measured $D_M/r_d$ and $D_H/r_d$ into constraints on the expansion history, reaching $5\%$ on $\Omegam$ and $1.0\%$ on $H_0$.
Pantheon+ alone constrains $\Omegam$ to $5.7\%$ but has no independent constraining power on $H_0$: with $M_B$ marginalized analytically, the sensitivity to the absolute distance scale is removed by construction and the $H_0$ posterior is set by the prior rather than by the data.
Direct late-Universe measurements fill this gap.
The SH0ES distance-ladder calibration combined with Pantheon+ gives $H_0 = 73.05\pm 1.04$~km/s/Mpc ($1.4\%$), with $\Omegam = 0.333\pm 0.018$ determined by the SNe.
H0LiCOW gives $H_0 = 73.33\pm 1.61$~km/s/Mpc ($2.2\%$) through strong-lensing time delays, with a weaker $\Omegam$ constraint because the sample contains only six lensed systems.
SNe~Ia and BAO constrain nearly orthogonal directions in the $H_0$--$\Omegam$ plane, and together with the CMB distance priors this complementarity is what makes the multi-probe joint analysis below markedly more precise than any single probe.

\label{sec:cobaya_validation}

\begin{figure}[htbp]
    \centering
    \includegraphics[width=0.48\textwidth]{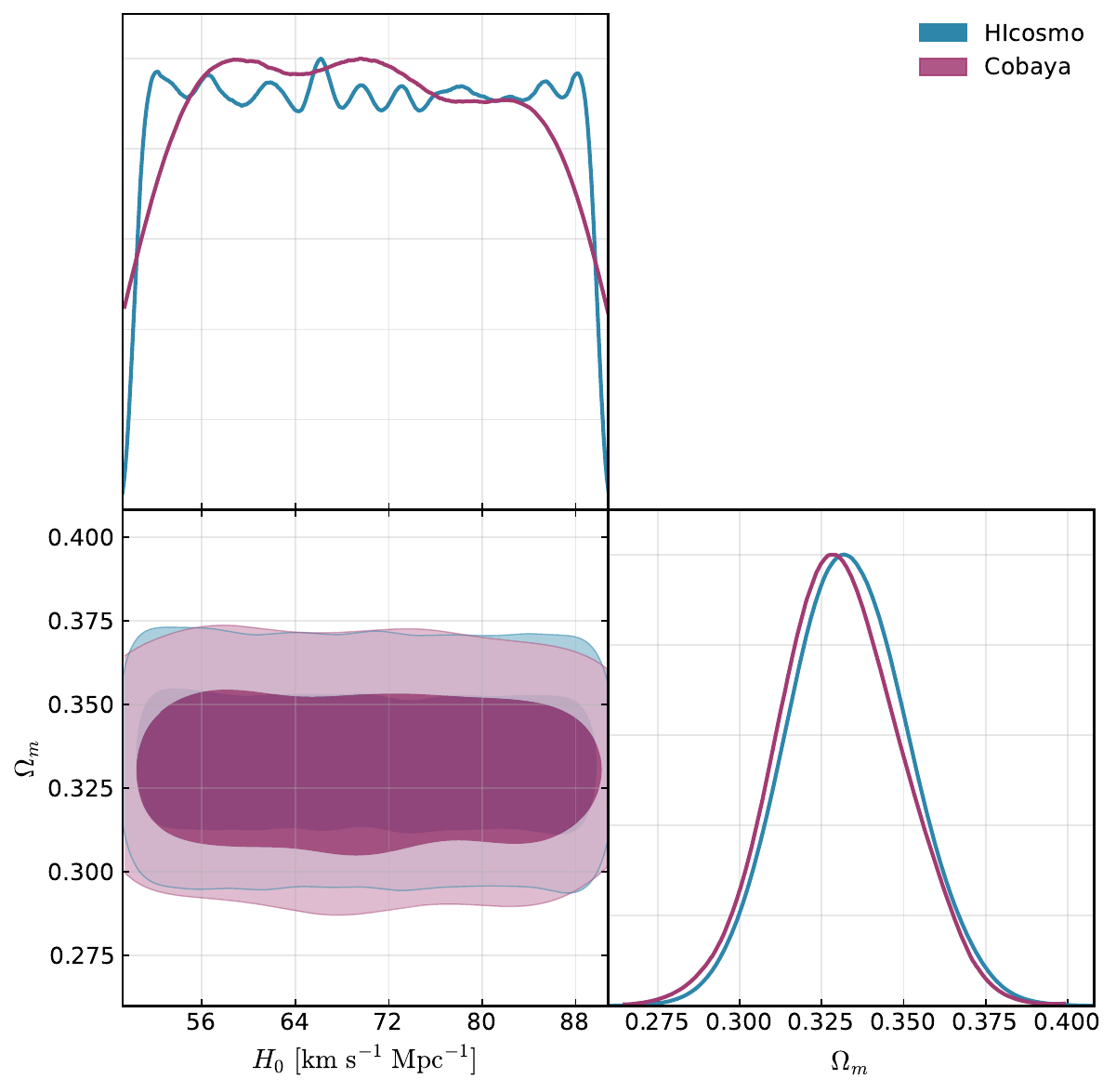}
    \hfill
    \includegraphics[width=0.48\textwidth]{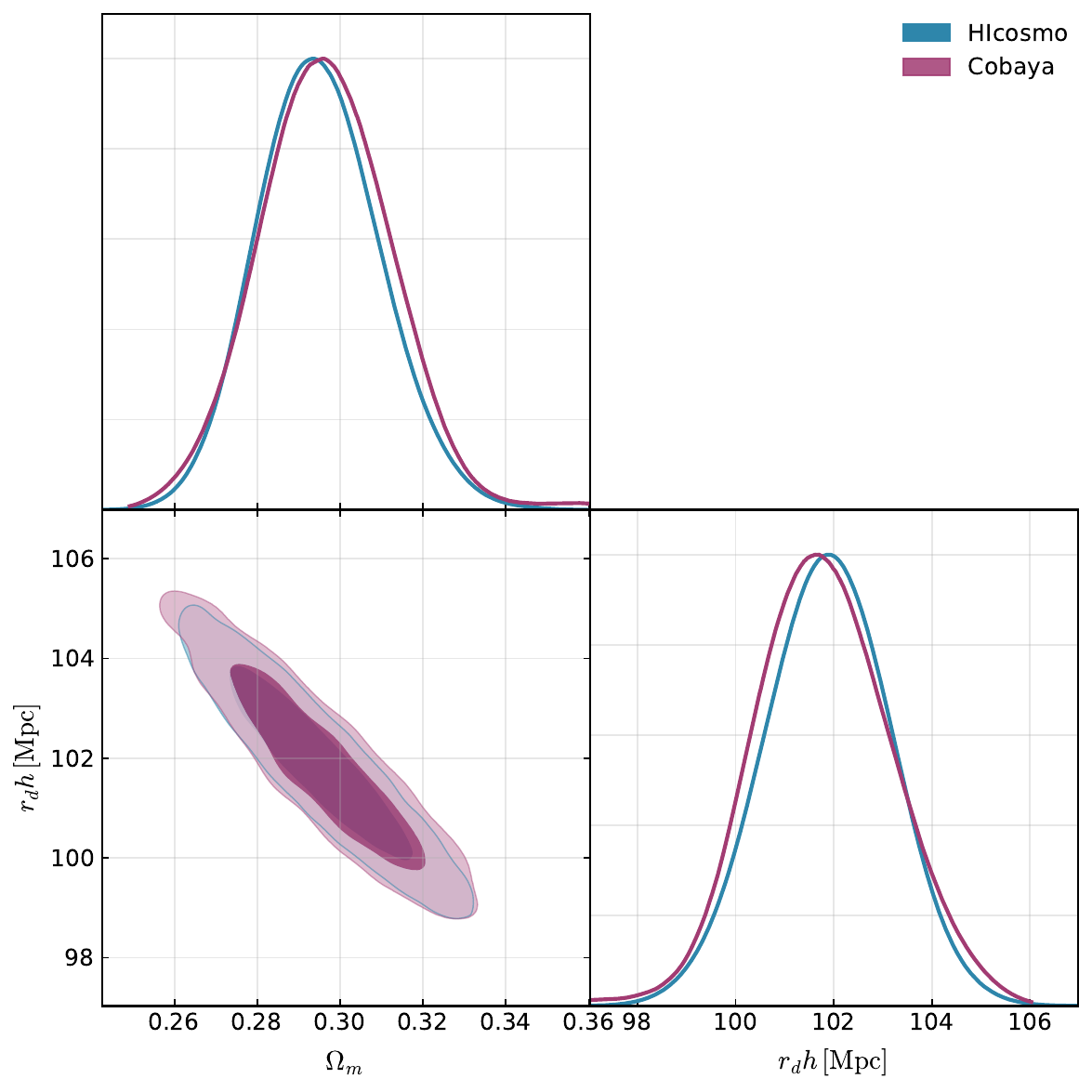}
    \\[6pt]
    \includegraphics[width=0.48\textwidth]{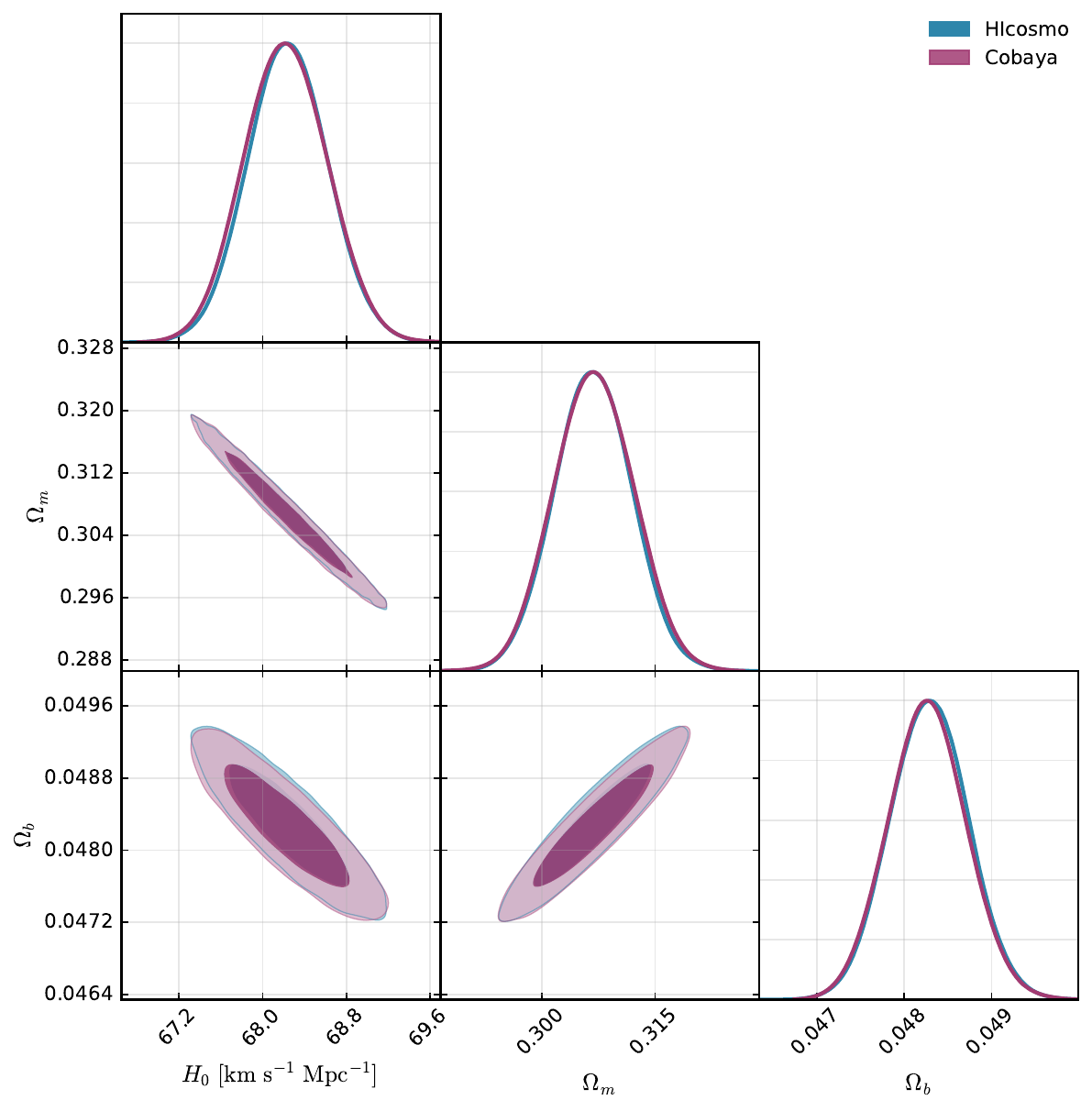}
    \caption{Comparison of \lcdm\ parameter constraints between \hicosmo\ and Cobaya.
    The two codes use identical data files.
    Upper left, Pantheon+ SNe Ia ($\Omegam$ deviation $< 0.2\sigma$, SN data alone cannot constrain $H_0$).
    Upper right, DESI 2024 BAO, compared in the $(\Omegam, r_d h)$ plane, the two parameters constrained by BAO alone \cite{DESI2024} (both agree within $0.1\sigma$).
    Bottom, joint SN+BAO+CMB (all parameters agree to better than $0.04\sigma$).
    Blue for \hicosmo, red for Cobaya.}
    \label{fig:cobaya}
\end{figure}

At the posterior level, we compare \hicosmo\ with Cobaya probe by probe (figure~\ref{fig:cobaya}).
Both codes fix $\Omegab h^2 = 0.02237$ (the Planck~2018 best-fit value), adopt the same massless-neutrino background ($N_{\rm eff}=3.046$), and read identical data files, so any difference at this level can only come from the inference itself.
With Pantheon+ alone, the $\Omegam$ constraints from the two codes agree within $0.2\sigma$.
With DESI BAO alone, we compare the posteriors in the $(\Omegam, r_d h)$ plane following the DESI convention \cite{DESI2024}; the two codes agree within $0.1\sigma$ on both parameters and jointly give $\Omegam = 0.295 \pm 0.015$ and $r_d h = (101.9 \pm 1.3)$~Mpc, consistent with the published DESI values.
In the joint SN+BAO+CMB analysis, \hicosmo\ evaluates $z_*$ and $r_s(z_*)$ with the CAMB-calibrated formulae of section~\ref{sec:architecture} while Cobaya calls CAMB itself, and the two codes agree on $H_0$, $\Omegam$, and $\Omegab$ to better than $0.04\sigma$.
Across all probe combinations the code-to-code differences stay between $0.04\sigma$ and $0.2\sigma$, well below the statistical uncertainties of current data.

The joint analysis of Pantheon+, DESI BAO, and Planck exploits the complementarity revealed by table~\ref{tab:single_probe} and tightens the $H_0$ constraint to the sub-percent level (bottom panel of figure~\ref{fig:cobaya}),
\begin{align}
    H_0 &= 68.24 \pm 0.38 \,\text{km/s/Mpc} \quad (0.6\%)\,, \\
    \Omegam &= 0.307 \pm 0.005 \quad (1.6\%)\,, \\
    \Omegab &= 0.0483 \pm 0.0004 \quad (0.8\%)\,.
\end{align}

\begin{figure}[htbp]
    \centering
    \includegraphics[width=0.85\textwidth]{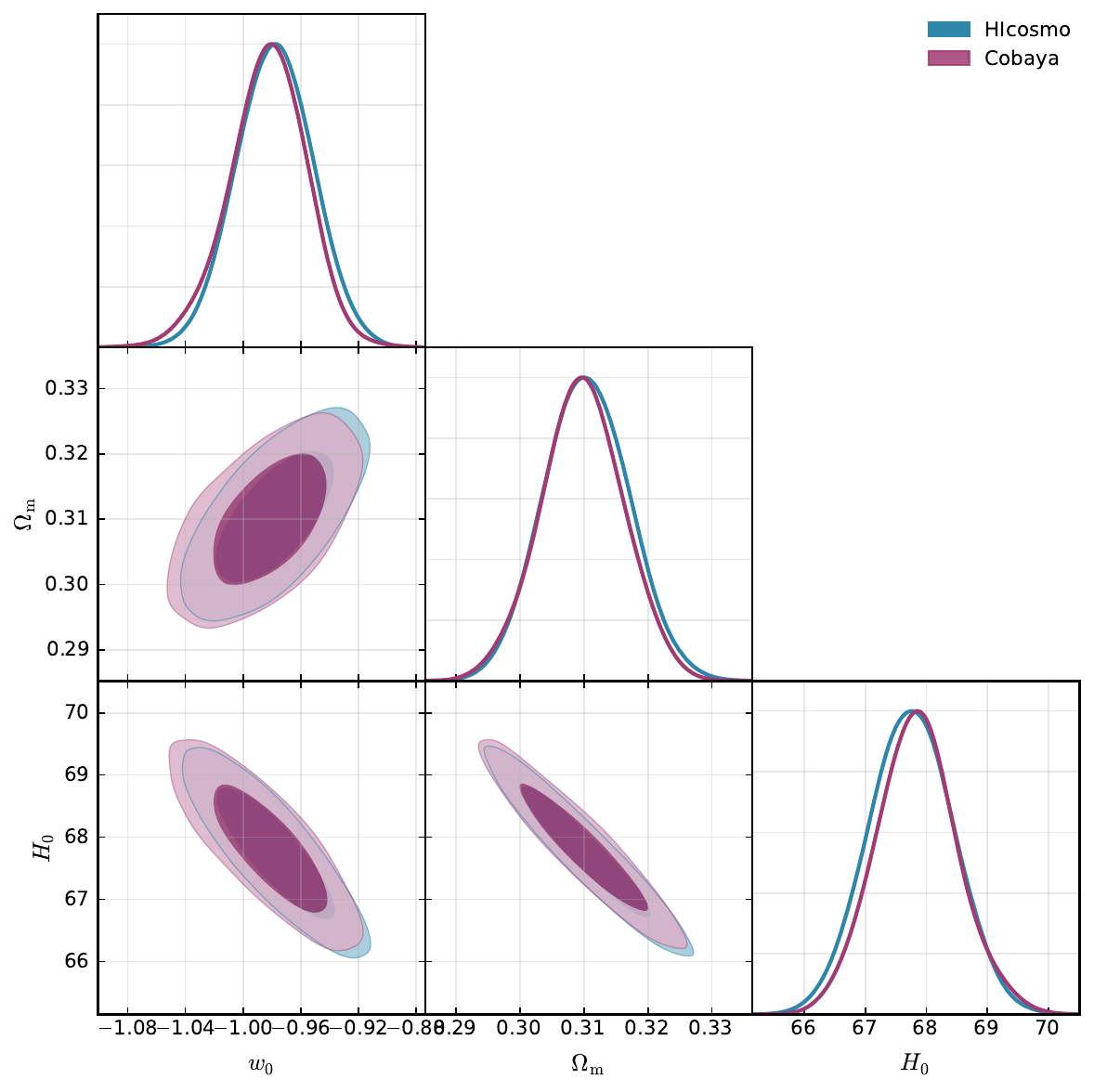}
    \caption{Comparison of \wcdm\ parameter constraints between \hicosmo\ and Cobaya (Pantheon+ + DESI DR1 BAO + Planck CMB distance priors).
    All parameters agree within $0.15\sigma$, showing that the \jax\ implementation reproduces the reference posterior in this extended background model, with deviations well below the statistical precision of the test.
    Blue for \hicosmo, red for Cobaya.}
    \label{fig:wcdm_cobaya}
\end{figure}

We next release the dark-energy equation of state.
For \wcdm\ with the same combined data, we find $w_0 = -0.978\pm 0.027$, consistent with the cosmological constant ($w_0=-1$) at the $0.8\sigma$ level. All posterior shifts between \hicosmo\ and Cobaya stay within $0.15\sigma$ in the $(w_0, \Omegam, H_0)$ space (figure~\ref{fig:wcdm_cobaya}).

\begin{figure}[htbp]
    \centering
    \includegraphics[width=0.85\textwidth]{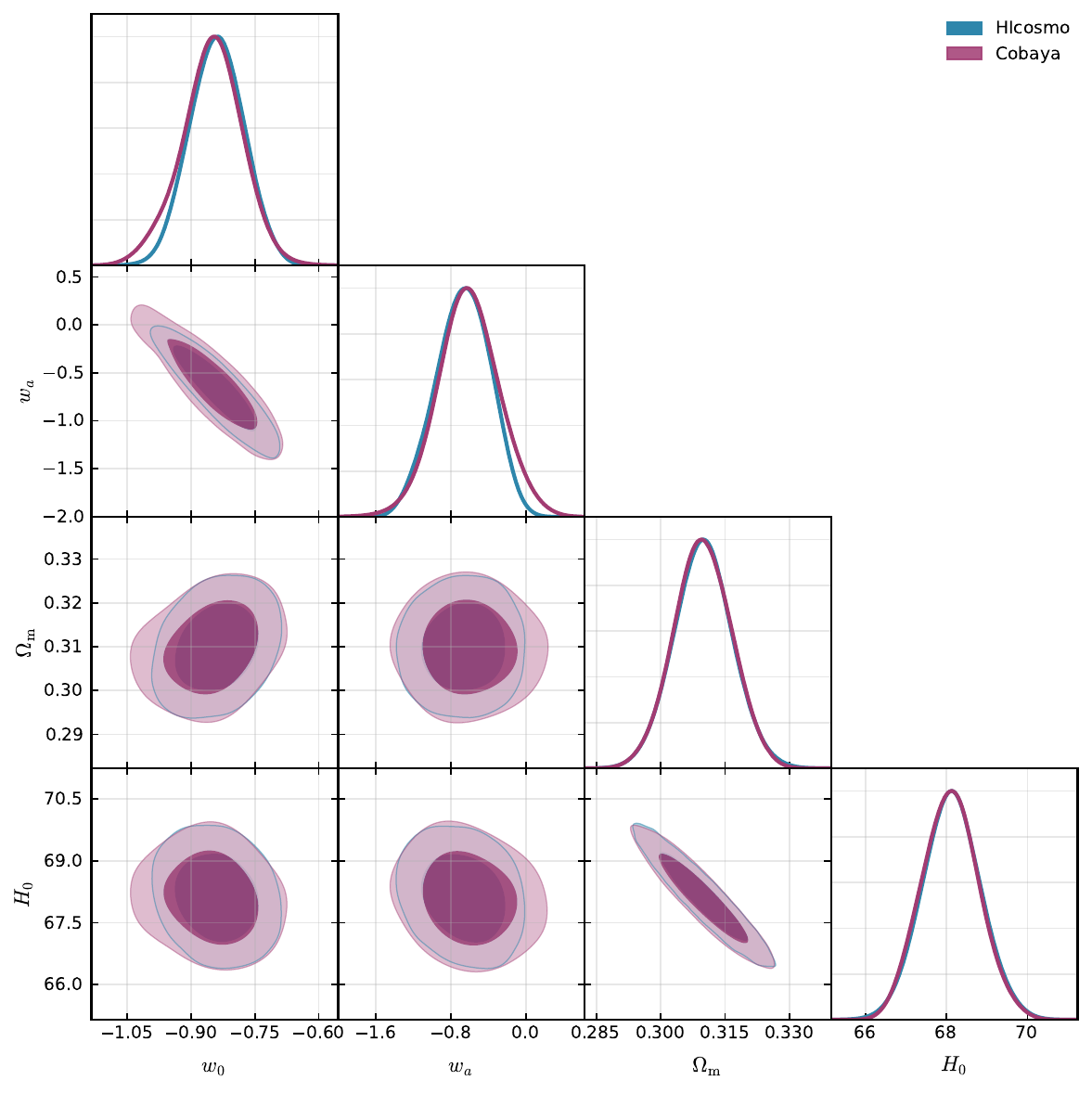}
    \caption{Comparison of $w_0 w_a$CDM $(w_0, w_a, \Omegam, H_0)$ constraints between \hicosmo\ and Cobaya (Pantheon+ + DESI DR1 BAO + Planck CMB distance priors).
    All parameters agree within $0.03\sigma$.
    The best fit has $w_0 > -1$ and $w_a < 0$, in the same direction as reported by DESI 2024 \cite{DESI2024}; note that the CMB distance priors used here are not equivalent to the full Planck likelihood.
    Blue for \hicosmo, red for Cobaya.}
    \label{fig:dark_energy}
\end{figure}

Figure~\ref{fig:dark_energy} shows the $w_0 w_a$CDM constraints in the full $(w_0, w_a, \Omegam, H_0)$ parameter space,
\begin{align}
    w_0 &= -0.84 \pm 0.06, \\
    w_a &= -0.67 \pm 0.29.
\end{align}
All code-to-code deviations remain within $0.03\sigma$.
The best-fit point deviates from the cosmological constant $(w_0, w_a)=(-1, 0)$ by about $2\sigma$, in the same direction as reported by DESI~2024 \cite{DESI2024}.
A quantitative comparison with the official DESI results requires caution. The DESI $w_0 w_a$CDM analysis uses the full Planck CMB likelihood, whereas the $(l_A, R, z_*)$ three-parameter distance priors used here retain considerably less information in the $w_0$--$w_a$ plane than the full likelihood (section~\ref{sec:architecture}).
The robustness of the deviation reported by DESI~2024 is itself still under discussion, with its significance depending on the details of the systematic-error treatment, the BAO reconstruction method, and the model-selection criteria \cite{DESI2024,Li:2026asg,Li:2025vuh,Li:2024qus,Du:2025xes}.

\begin{figure}[htbp]
    \centering
    \includegraphics[width=0.6\textwidth]{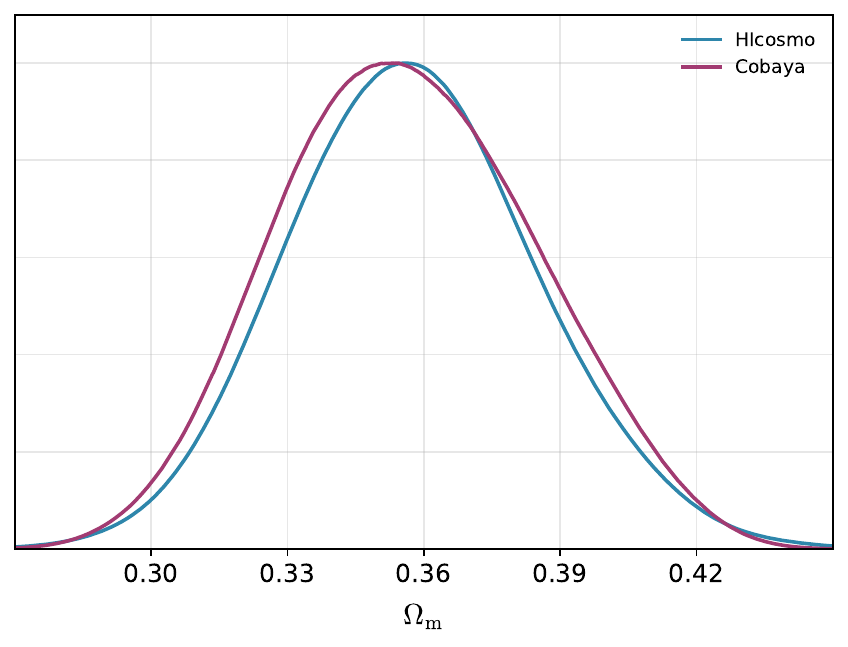}
    \caption{Comparison of the $\Omegam$ posterior between \hicosmo\ and Cobaya for the Union3 Type Ia supernovae \cite{Rubin2023} (2087 supernovae, 22 binned distance moduli).
    The two codes use identical data files and agree within $0.2\sigma$.
    SN data alone cannot constrain $H_0$ (the absolute magnitude is marginalized).
    Blue for \hicosmo, red for Cobaya.}
    \label{fig:cobaya_new}
\end{figure}

The comparison is further extended to the Union3 supernova compilation \cite{Rubin2023}, which is standardized with the UNITY1.5 Bayesian hierarchical framework rather than the traditional Tripp formula.
As shown in figure~\ref{fig:cobaya_new}, the $\Omegam$ constraints from the two codes again agree within $0.2\sigma$.
Across \lcdm, \wcdm, and $w_0 w_a$CDM, and across Pantheon+, Union3, DESI BAO, and Planck data, the posterior shifts between \hicosmo\ and Cobaya remain within $0.2\sigma$, demonstrating that the two independent implementations yield statistically indistinguishable inference results.

\section{Performance benchmarks}
\label{sec:performance}

Section~\ref{sec:validation} confirmed the numerical agreement between \hicosmo\ and the reference codes. This section examines the computational cost of achieving that agreement.
We measure three quantities, namely the single-evaluation time of the likelihood, the end-to-end efficiency of full MCMC sampling, and the cost of the Fisher-matrix computation on the same likelihood.
Single-evaluation timings are measured on an Apple~M1~Max (10-core CPU, 32\,GB RAM). The MCMC sampling efficiency (table~\ref{tab:ess_comparison}) and the GPU scaling benchmark (table~\ref{tab:gpu_benchmark}) are measured on a server with an Intel Xeon w9-3495X CPU (8 cores) and an NVIDIA RTX~A6000 (48\,GB) GPU.
All timings are averages over $10^3$--$10^4$ evaluations after JIT warmup and therefore correspond to the steady-state cost of a real MCMC run.

As described in section~\ref{sec:architecture}, the first call of a \jax\ function triggers the XLA compilation, and subsequent calls with the same input shapes execute the cached compiled program.
The one-time compilation cost of the three core computations ranges from 73 to 310~ms. The post-compilation steady-state evaluation times are 0.05~ms for the distance calculation, 0.95~ms for the SN likelihood, and 0.18~ms for the BAO likelihood.
For an MCMC run with $10^4$--$10^6$ likelihood evaluations, the compilation overhead is amortized over the whole run and the cold-start latency is negligible.

\begin{figure}[htbp]
    \centering
    \includegraphics[width=0.85\textwidth]{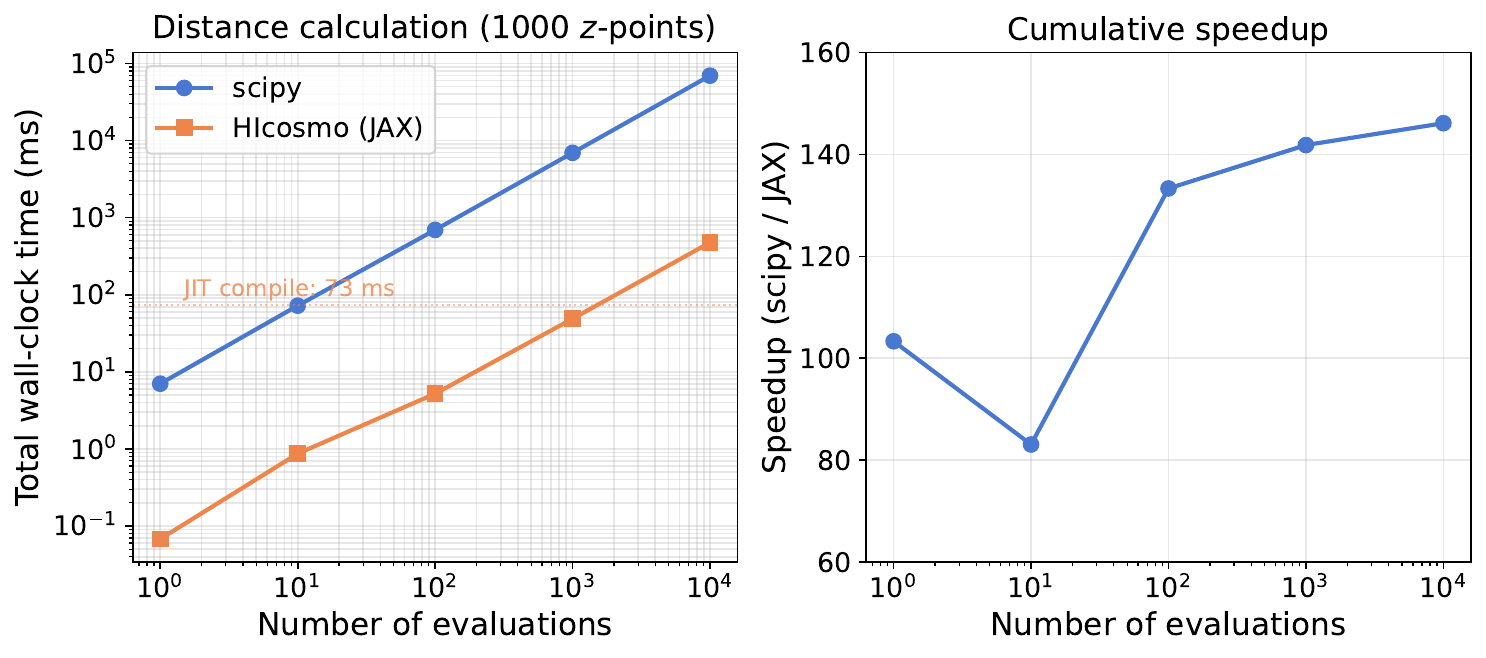}
    \caption{Performance benchmark of \hicosmo\ against \texttt{scipy} for the distance calculation (1000 $z$ points).
    Left, total wall-clock time as a function of the number of repeated evaluations $N$ (log--log scale).
    Both methods grow linearly, but after the one-time JIT compilation (dashed line, $\sim$50~ms) the per-call cost of \hicosmo\ (orange) is lower by a factor of about 140.
    Right, cumulative speedup versus $N$, showing that the post-compilation speedup is nearly constant over all evaluation counts.}
    \label{fig:benchmark}
\end{figure}

Figure~\ref{fig:benchmark} quantifies the comparison between \hicosmo\ and a \texttt{scipy} implementation.
The left panel shows the total wall-clock time of the distance calculation (comoving distances at 1000 redshift points per call) repeated $N$ times.
After the JIT compilation ($\sim$50~ms), each \hicosmo\ call takes about 0.05~ms versus about 7~ms for \texttt{scipy}, and the speedup remains stable at about $140\times$ from $N=1$ to $N=10{,}000$.
A simplified SN-likelihood benchmark in the same figure yields a speedup of about $8\times$.
The right panel shows that the speedup is nearly constant in $N$, consistent with the warm phase being dominated by the compiled numerical kernels rather than by the Python interpreter.
This comparison is not made at strictly matched numerical precision. The \texttt{scipy} adaptive quadrature uses a tolerance of $10^{-8}$, while the fixed grid of \hicosmo\ reaches about $2\times 10^{-7}$.
Both errors are far below typical posterior uncertainties, so the comparison should be read as a speedup at equivalent parameter-estimation precision.

On the single-evaluation matched pipeline, including the distance integral, the analytic $M_B$ marginalization, and the full-covariance $\chi^2$, \hicosmo\ takes 1.49~ms per call on the CPU versus 23.67~ms for \texttt{scipy}, a speedup of about $16\times$.
At the Pantheon+ size, however, the GPU offers no advantage on a single evaluation, as the $1580\times 1580$ linear algebra is too small to fill the device, and per-call dispatch and host--device synchronization dominate.
For inference the relevant quantity is the throughput of a full sampling run, where the warmup and sampling loops are compiled into a single on-device program and the per-call dispatch overhead is amortized.
Since the identical \jax\ code runs on both devices, we ask how much the GPU accelerates \hicosmo\ over its own CPU as the dataset grows.
Table~\ref{tab:gpu_benchmark} reports this for an \lcdm\ supernova constraint, with the dataset size $N_{\rm SN}$ growing from the Pantheon+ value to $20{,}000$.
At the Pantheon+ size the two devices are comparable, but as $N_{\rm SN}$ grows the GPU pulls ahead, reaching a $20.8\times$ speedup at $N_{\rm SN}=20{,}000$.

The speedup grows because each NUTS gradient is dominated by the $\mathcal{O}(N_{\rm SN}^2)$ covariance matrix--vector product, a cost the CPU pays serially while the GPU evaluates it in parallel. At $N_{\rm SN}=20{,}000$ a full run takes 4.3~hours on the CPU but only 13~minutes on the GPU.
The same \jax\ code runs on the GPU without modification. When facing massive survey datasets in the future, the data can be processed efficiently on the GPU with no code changes required.
Whether similar gains apply to BAO, weak-lensing, or perturbation-level likelihoods depends on the numerical structure of each problem and requires separate benchmarks.

\begin{table}[htbp]
\centering
\caption{ESS throughput of \hicosmo\ on the CPU versus the GPU as the dataset size $N_{\rm SN}$ grows, using synthetic SN data with a dense $N_{\rm SN}\times N_{\rm SN}$ covariance matrix and the identical \jax\ code on both devices. Intel Xeon w9-3495X (CPU) and NVIDIA RTX~A6000 (GPU).}
\label{tab:gpu_benchmark}
\begin{tabular}{lccc}
\toprule
$N_{\rm SN}$ & \hicosmo\ CPU (ESS/s) & \hicosmo\ GPU (ESS/s) & GPU speedup \\
\midrule
$1580$ & $265$ & $310$ & $1.2\times$ \\
$5{,}000$ & $12.6$ & $119$ & $9.4\times$ \\
$20{,}000$ & $0.5$ & $10.4$ & $20.8\times$ \\
\bottomrule
\end{tabular}
\end{table}

The speedup of a single likelihood evaluation does not by itself determine the cost of inference.
The end-to-end sampling efficiency is characterized by the effective sample size, $\mathrm{ESS} = N/\tau$, where $N$ is the total number of draws and $\tau$ the integrated autocorrelation time.
If the sampler produces highly correlated samples, even a fast likelihood can yield fewer independent samples per unit time than a slow likelihood paired with an efficient sampler.

For a quantitative comparison, we analyze the same \lcdm\ + Pantheon+ problem with \hicosmo\ and Cobaya \citep{Cobaya2021} under a matched 8-core, 8-chain configuration.
Both analyses apply the standard Pantheon+ cut $z_{\rm HD}>0.01$ to the 1580 SNe~Ia \cite{Scolnic2022}, marginalize $M_B$ analytically, and adopt flat priors on $H_0$ and $\Omegam$.
\hicosmo\ runs 8 NUTS chains, and Cobaya runs 8 Metropolis--Hastings chains with single-threaded CAMB on the same 8 cores.

Table~\ref{tab:ess_comparison} summarizes the results.
The two samplers differ mainly in sample autocorrelation.
The NUTS trajectories of \hicosmo\ are nearly independent ($\tau\approx 1.2$), whereas the Metropolis--Hastings chains of Cobaya have $\tau\approx 11.4$, so each \hicosmo\ draw contains roughly $9\times$ as much independent information.
Together with the JIT-compiled likelihood, this gives an ESS throughput of $400$~ESS/s on the CPU, about $8.7\times$ that of Cobaya ($45.8$~ESS/s).
At this dataset size the CPU outperforms the GPU ($291$~ESS/s) because the $1580\times 1580$ linear algebra is too small to fill the device, and per-call dispatch overhead dominates.
The GPU pays off only at survey scale, where it accelerates \hicosmo\ by up to $20\times$ over its own CPU execution as $N_{\rm SN}$ grows (table~\ref{tab:gpu_benchmark}).

\begin{table}[htbp]
\centering
\caption{Effective-sample-size (ESS) throughput on the real Pantheon+ data (1580 SNe~Ia), 8 cores, 8 chains. ESS/s is the number of independent posterior samples per second. Intel Xeon w9-3495X (CPU) and NVIDIA RTX~A6000 (GPU).}
\label{tab:ess_comparison}
\begin{tabular}{lcccc}
\toprule
Method & Wall (s) & $\tau$ & ESS/s & vs.\ Cobaya \\
\midrule
Cobaya (MH, 8 chains) & $15.6$ & $11.4$ & $45.8$ & --- \\
\hicosmo\ (NUTS, 8 chains, CPU) & $21.5$ & $1.2$ & $400$ & $8.7\times$ \\
\hicosmo\ (NUTS, GPU) & $27.4$ & $1.2$ & $291$ & $6.4\times$ \\
\bottomrule
\end{tabular}
\end{table}

The third benchmark is the Fisher-matrix computation.
At the reference point of the Pantheon+ SN likelihood ($H_0=70$~km/s/Mpc, $\Omegam=0.3$), the AD Hessian yields $\sigma(\Omegam)=0.01703$ in 4.7~ms, roughly the cost of eight likelihood evaluations, as expected for nested reverse-mode differentiation \cite{Griewank2008}, and with no step size to choose at any stage.
Central finite differences on the same likelihood are stable over a wide range of step sizes, $h\sim 10^{-2}$--$10^{-6}$, with deviations below $0.05\%$ (table~\ref{tab:fisher_precision}).
For $h\leq 10^{-7}$, however, floating-point round-off dominates, with deviations of $0.2\%$ at $h=10^{-7}$ and $46\%$ at $h=10^{-8}$, the classic failure mode of finite-difference Hessians \cite{Griewank2008}.
Automatic differentiation differentiates the computation of the likelihood itself rather than probing it with perturbations, so the round-off failure mode does not arise and no step size needs to be chosen.

\begin{table}[htbp]
\centering
\caption{Fisher-matrix precision, AD versus central finite differences for the Pantheon+ SN likelihood ($\Omegam$--$H_0$ plane, reference point $H_0=70$, $\Omegam=0.3$, Apple~M1~Max).
AD differentiates the computation of the likelihood directly and requires no step size.
Finite differences with a uniform step $h$ in both parameters are stable over a wide range but break down when floating-point round-off dominates at small $h$.}
\label{tab:fisher_precision}
\begin{tabular}{lccc}
\toprule
Method & $\sigma(\Omegam)$ & Deviation from AD & Time \\
\midrule
AD (automatic differentiation) & 0.01703 & --- & 4.7 ms \\
\midrule
FD ($h = 10^{-2}$) & 0.01702 & 0.04\% & 9 ms \\
FD ($h = 10^{-3}$) & 0.01703 & ${<}\,0.01\%$ & 9 ms \\
FD ($h = 10^{-4}$) & 0.01703 & ${<}\,0.01\%$ & 9 ms \\
FD ($h = 10^{-5}$) & 0.01703 & ${<}\,0.01\%$ & 9 ms \\
FD ($h = 10^{-6}$) & 0.01703 & 0.01\% & 9 ms \\
FD ($h = 10^{-7}$) & 0.01707 & 0.21\% & 9 ms \\
FD ($h = 10^{-8}$) & 0.00926 & 46\% & 9 ms \\
\bottomrule
\end{tabular}
\end{table}

\section{Fisher forecasts for 21\,cm intensity-mapping surveys}
\label{sec:applications}
\label{sec:fisher_21cm}

The applications so far are MCMC analyses. The same differentiable likelihood also supports Fisher forecasting, which assesses the constraining power of future surveys without running full chains.
Because the Hessian comes from automatic differentiation rather than finite differences, the forecasts do not depend on any step-size choice (table~\ref{tab:fisher_precision}).
We apply this method to 21\,cm intensity-mapping surveys, which trace large-scale structure through the power spectrum of neutral-hydrogen 21\,cm emission.

The forecasting method follows the standard Fisher framework \cite{Bull:2014rha,Bull:2015lja,Zhang:2019ipd,Zhang:2019dyq,Zhang:2021yof,Wu:2021vfz,Wu:2022dgy,Zhang:2023gaz,Wu:2022jkf,Pan:2024xoj}.
The signal is the anisotropic 21\,cm power spectrum,
\begin{equation}
    P_{21}(k,\mu,z) = \bar{T}_b^2(z)\, b_{\rm HI}^2(z)\, \left(1+\beta\mu^2\right)^2 P_m(k,z),
\end{equation}
where the mean brightness temperature $\bar{T}_b(z)$ is determined by the neutral-hydrogen abundance $\Omega_{\rm HI}(z)$ and the expansion history, $\beta = f/b_{\rm HI}$ is the redshift-space-distortion parameter, and $\mu$ is the cosine of the angle between the wavevector and the line of sight.
The noise power is set by the system temperature (including the sky contribution), the number of receiving elements, and the total integration time. The beam window function suppresses transverse modes below the angular resolution, and the minimum accessible wavenumber $k_{\rm min}$ in each redshift bin is set by the survey volume of that bin.
Nonlinear scales are excluded with the redshift-dependent cutoff $k_{\rm max}(z) = 0.14\,(1+z)^{2/(2+n_s)}\,\mathrm{Mpc}^{-1}$ adopted in ref.~\cite{Bull:2014rha}.
Following ref.~\cite{Bull:2014rha}, the HI bias in each redshift bin is marginalized as a nuisance parameter, and the resulting Fisher matrix is projected onto the \wcdm\ parameters $(w, \Omegam, H_0)$.
We do not model foreground removal explicitly, so these forecasts should be regarded as optimistic estimates for the given survey configurations.
Table~\ref{tab:fisher_surveys} lists the forecast $1\sigma$ constraints on the equation-of-state parameter $w$ (marginalized over the remaining parameters) for six surveys under construction or in operation, each characterized by its frequency range, redshift coverage, and sky area.

In \hicosmo, a complete Fisher forecast for a given survey requires only a few lines of code.
\begin{lstlisting}
from hicosmo.fisher import load_survey, IntensityMappingFisher
from hicosmo.models import wCDM

survey = load_survey('ska1_wide_band1')
fisher = IntensityMappingFisher(survey, wCDM())
result = fisher.parameter_forecast(
    ['w0'], marginalize_over=['H0', 'Omega_m'])
\end{lstlisting}
\texttt{load\_survey} reads the survey parameters from a YAML configuration file, and \texttt{parameter\_forecast} computes the power-spectrum Fisher matrix, marginalizes over nuisance parameters, and projects onto the target parameters. Adding a new survey requires only a new YAML file, with no code changes.

\begin{table}[htbp]
\centering
\caption{Fisher forecasts for 21\,cm intensity-mapping surveys.
Listed are the forecast $1\sigma$ constraints on the dark-energy parameter $w$ in the \wcdm\ model.
The fiducial cosmology is the Planck 2018 best fit.
The computation includes redshift-space distortions, instrumental noise, and the beam, with the HI bias of each redshift bin marginalized. Foreground removal is not modeled explicitly.}
\label{tab:fisher_surveys}
\begin{tabular}{lcccc}
\toprule
Survey & Frequency Band & $z$ Range & Sky Area & $\sigma(w)$ \\
\midrule
SKA1 Band 1 & 350--1050 MHz & 0.30--3.00 & 20k deg$^2$ & 0.046 \\
SKA1 Band 2 & 950--1420 MHz & 0.00--0.50 & 5k deg$^2$ & 0.209 \\
MeerKAT & 580--1420 MHz & 0.22--0.97 & 8k deg$^2$ & 0.125 \\
BINGO & 960--1260 MHz & 0.15--0.45 & 4k deg$^2$ & 0.292 \\
Tianlai & 550--950 MHz & 0.50--1.55 & 10k deg$^2$ & 0.091 \\
CHIME & 400--800 MHz & 0.80--2.50 & 22.6k deg$^2$ & 0.061 \\
\bottomrule
\end{tabular}
\end{table}

The differences in constraining power in table~\ref{tab:fisher_surveys} mainly reflect the instrumental configuration and observing strategy of each survey.
SKA1 Band~1 covers $0.30<z<3.00$ over $20{,}000$~deg$^2$, and its long redshift baseline breaks the $w$--$\Omegam$ degeneracy, yielding $\sigma(w)=0.046$.
SKA1 Band~2 covers only the nearby Universe ($z<0.5$) and is correspondingly weaker ($\sigma(w)=0.209$), with its main value lying in the low-redshift information it contributes in combination with Band~1.
CHIME \cite{CHIME2022} consists of four $100\,\mathrm{m}\times 20\,\mathrm{m}$ cylindrical reflectors with a total of 1024 feeds; its drift-scan strategy covers $22{,}600$~deg$^2$, and three years of integration time combined with the wide redshift baseline ($0.80<z<2.50$) yield $\sigma(w)=0.061$, the strongest constraint among current-generation surveys.
The full-scale Tianlai array (eight $120\,\mathrm{m}\times 15\,\mathrm{m}$ cylinders with 2048 feeds \cite{Xu:2014bya}) covers $0.50<z<1.55$ and yields $\sigma(w)=0.091$.
MeerKAT has a smaller sky area but reaches intermediate redshifts, giving $\sigma(w)=0.125$.
BINGO is purpose-built for BAO in a narrow low-redshift window and gives the loosest forecast, $\sigma(w)=0.292$.

\begin{figure}[htbp]
    \centering
    \includegraphics[width=0.75\textwidth]{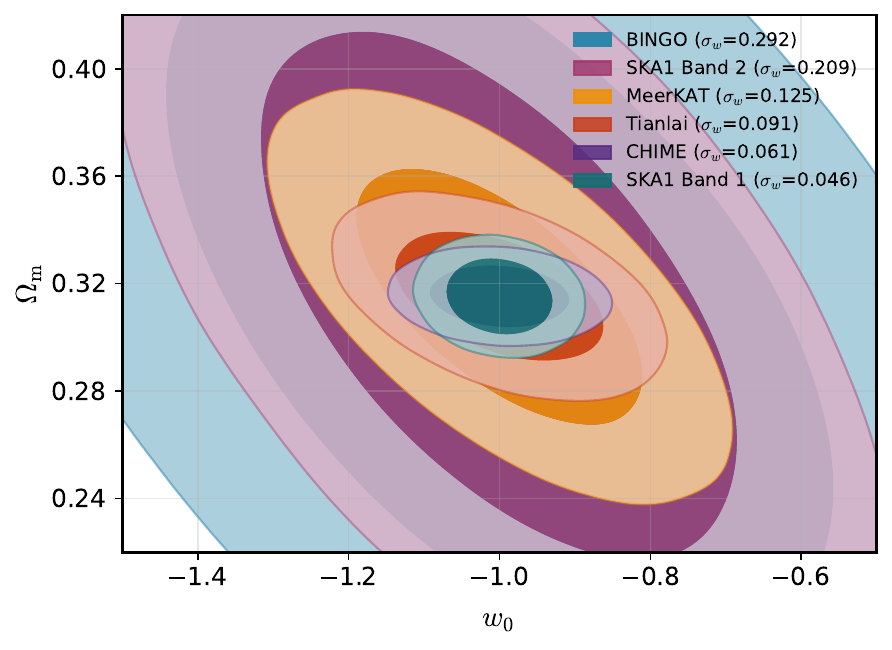}
    \caption{Fisher forecasts for the dark-energy equation of state $w_0$ and the matter density $\Omegam$ from six 21\,cm intensity-mapping surveys.
    Filled ellipses show the $1\sigma$ ($2\sigma$) confidence regions of each survey (marginalized over $H_0$).
    The cross marks the \lcdm\ fiducial point ($w_0 = -1$, $\Omegam = 0.3153$).}
    \label{fig:fisher_21cm}
\end{figure}

Figure~\ref{fig:fisher_21cm} shows the Fisher forecast ellipses in the $w_0$--$\Omegam$ plane for each survey. The surveys differ not only in absolute precision but also in the orientation of their degeneracies: the ellipses of SKA1 Band~1 and Tianlai are nearly parallel (both extend to $z>1$), while the major axis of CHIME is slightly rotated. This complementarity in degeneracy directions indicates that a joint analysis has the potential to tighten the constraints further.

\begin{figure}[htbp]
    \centering
    \includegraphics[width=0.88\textwidth]{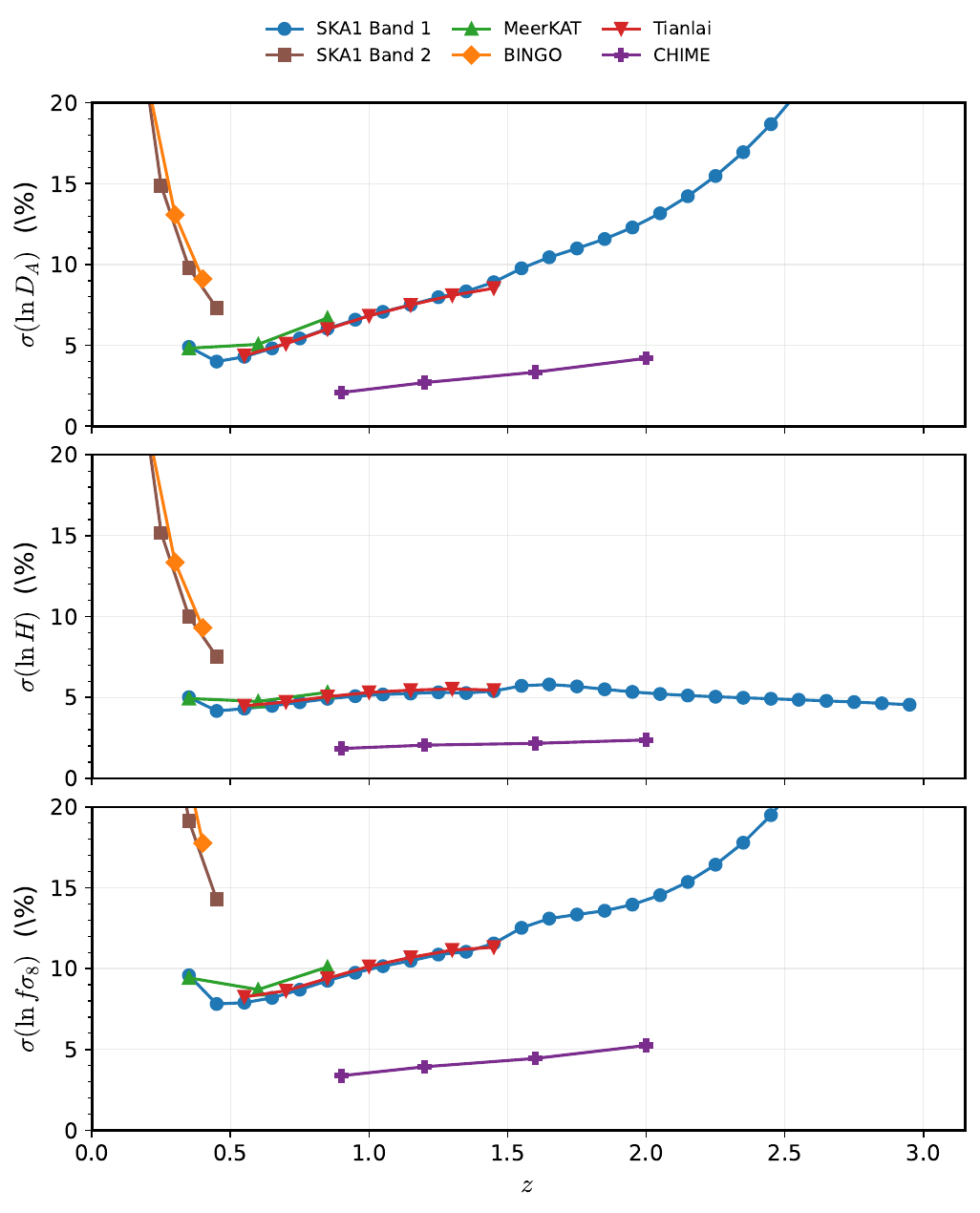}
    \caption{Per-bin Fisher precision for six 21\,cm intensity-mapping surveys on three key observables, namely angular diameter distance $\sigma(\ln D_A)$ (top), Hubble parameter $\sigma(\ln H)$ (middle), and growth rate $\sigma(\ln f\sigma_8)$ (bottom).
    The vertical axis shows percentage precision, truncated at 20\% (larger values indicate bins carrying negligible information).
    The HI bias in each bin has been marginalized as a nuisance parameter.}
    \label{fig:fisher_observables}
\end{figure}

Before projecting onto cosmological parameters, the Fisher matrix constrains four observables per redshift bin, namely the angular diameter distance $\ln D_A$, the Hubble parameter $\ln H$, the growth rate times the clustering amplitude $\ln(f\sigma_8)$, and the HI bias $\ln(b\sigma_8)$.
Figure~\ref{fig:fisher_observables} displays the per-bin precision for the three key observables across all six surveys.
CHIME, with its 1024 feeds, maintains $\sigma(\ln D_A)\approx 2$--$4\%$ and $\sigma(\ln H)\approx 2\%$ across all redshift bins.
SKA1 Band~1 reaches $\sigma(\ln D_A)\approx 4$--$8\%$ and $\sigma(\ln H)\approx 4$--$5\%$ at $z<1.5$, but its distance precision degrades rapidly at $z>2.5$, reflecting the combined effect of decreasing brightness temperature and rising foreground noise at high redshift; nevertheless, the total mode count from 26 bins spanning $0.35<z<3.0$ gives SKA1 Band~1 the tightest $\sigma(w)$.
Tianlai and MeerKAT achieve comparable per-bin precision within their respective intermediate-redshift windows.
The growth-rate precision $\sigma(\ln f\sigma_8)$ follows a similar pattern but is systematically lower than the distance observables, because the Fisher information on redshift-space distortions enters only through the $\mu$-dependence, whereas the distance information draws on the total mode count.

\section{Discussion and conclusions}
\label{sec:discussion}
\label{sec:conclusion}

The results of the previous sections show that, for background-level observables (SN, BAO, CMB distance priors), a single differentiable code path can reproduce established constraints, reduce the cost of repeated likelihood evaluations, and supply the derivatives required by NUTS sampling and Fisher forecasting.
This conclusion applies to background cosmology.
Background likelihoods are fast to evaluate and operate on highly compressed data, and they are not the main computational bottleneck of Stage~IV cosmology.
We chose the background level because mature reference computations and published constraints are available here for strict comparison, so that every step of the differentiable design can be checked, and the outcome of those checks takes the form of a directly usable tool.
This section places \hicosmo\ in the landscape of existing cosmological inference codes, discusses its limitations, and analyzes what is required to extend the same design to the perturbation level.

\subsection{Comparison with existing tools}

\begin{table}[htbp]
\centering
\small
\caption{Comparison of cosmological parameter-estimation tools. The Scope column indicates the physics level covered, namely background distances (Bkg), CMB/$P(k)$ perturbations (Pert), or weak-lensing/galaxy-clustering (WL/LSS).}
\label{tab:comparison}
\begin{tabular}{lccccccc}
\toprule
Tool & Backend & Sampler & JIT & AD & GPU & Boltzmann & Scope \\
\midrule
CosmoMC & Fortran & MH & -- & -- & -- & CAMB & Pert \\
Cobaya & Python & MH/MCMC & -- & -- & -- & CAMB/CLASS & Pert \\
MontePython & Python & MH/NS & -- & -- & -- & CLASS & Pert \\
emcee & Python & Ensemble & -- & -- & -- & -- & General \\
\midrule
jax-cosmo & JAX & external & \checkmark & \checkmark & \checkmark & emulator & WL/LSS \\
CosmoPower & JAX & external & \checkmark & \checkmark & \checkmark & NN emulator & Pert \\
\hicosmo & JAX & NUTS & \checkmark & \checkmark & \checkmark & -- & Bkg \\
\bottomrule
\end{tabular}
\end{table}

The performance gains of \hicosmo\ depend on the task and cannot be summarized by a single speedup factor.
For distance calculations, JIT compilation and vectorization reduce one evaluation at 1000 redshift points from the millisecond level to $0.05$~ms.
The matched-pipeline Pantheon+ likelihood ($N_{\rm SN}=1580$) gains $15.9\times$ on the CPU over a \texttt{scipy} baseline.
On the GPU, \hicosmo\ accelerates over its own CPU execution by a factor that grows with $N_{\rm SN}$, reaching about $20\times$ at $N_{\rm SN}=20{,}000$.
These figures show that the speedup varies with the type of computation and depends on how well the arithmetic intensity matches the hardware parallelism.

The tools listed in table~\ref{tab:comparison} target different use cases, and the comparison must respect their respective positioning.
CosmoMC is the de facto standard for CMB power-spectrum analyses, deeply integrated with CAMB and very mature in its coverage, but extending the Fortran codebase with new likelihoods or samplers is difficult.
Cobaya offers excellent modularity and multi-sampler support in Python, but its core numerics rest on \texttt{scipy} and gradients are not readily available.
MontePython is tightly integrated with CLASS and has a rich likelihood library, but its main configuration does not support gradient-based sampling.
emcee is a widely used general-purpose ensemble sampler with no cosmology-specific functionality.
Among the \jax-native tools, jax-cosmo \citep{JAXCosmo} provides differentiable weak-lensing and galaxy-clustering pipelines, usually combined with external samplers such as NumPyro or BlackJAX.
CosmoPower \citep{CosmoPower,Piras2024} replaces the Boltzmann solver with neural-network emulators of $C_\ell$ and $P(k)$, whose forward pass and gradients have been ported to \jax.
\hicosmo\ does not overlap with any of the above. jax-cosmo and CosmoPower target perturbation-level observables, whereas \hicosmo\ targets background-level likelihoods (SN, BAO, CMB distance priors, strong lensing), with a built-in NUTS sampler coupled directly to the differentiable likelihoods.
This focus brings concrete returns.
On the real Pantheon+ \lcdm\ benchmark the CPU ESS throughput is about $8.7\times$ that of Cobaya under a matched 8-core configuration, and the GPU accelerates \hicosmo\ by up to $20\times$ over its own CPU at survey scale.
The corresponding cost is that \hicosmo\ contains no Boltzmann solver, and any perturbation-level analysis must obtain $C_\ell$ or $P(k)$ from external tools.

\subsection{Limitations and scope of applicability}

\hicosmo\ is designed for background-cosmology parameter estimation, including constraints on $H_0$, $\Omegam$, $\Omegab$, $\Omegak$, and the dark-energy equation-of-state parameters ($w_0$, $w_a$), multi-probe joint analyses, and Fisher forecasts.
Perturbation-level computations, such as full CMB power-spectrum analyses, $P(k)$ constraints, and scale-dependent growth factors, require a Boltzmann solver and are outside the scope of the framework and of the validation presented in this paper.
Users who need such analyses should turn to CAMB \cite{CAMB}, CLASS \cite{CLASS}, or differentiable emulators such as CosmoPower-JAX \cite{CosmoPower,Piras2024}, and treat \hicosmo\ as the background layer of a larger pipeline.

The numerical approximations common to all analyses, including the distance-integration truncation ($<2\times10^{-7}$) and floating-point round-off, are negligible compared with the statistical uncertainties of the posteriors.
The most delicate derived quantity is the sound horizon $r_d$.
Because BAO constraints approximately satisfy $H_0\propto r_d^{-1}$, the $1$--$2\%$ offset of the raw Eisenstein--Hu formula would translate into $0.7$--$1.4$~km/s/Mpc at $H_0\approx 68$~km/s/Mpc, larger than the $1\sigma$ uncertainty of the joint constraint in section~\ref{sec:applications} ($0.38$~km/s/Mpc).
For this reason \hicosmo\ evaluates $r_d$ with the CAMB-calibrated Eisenstein--Hu formula throughout (section~\ref{sec:architecture}), which agrees with the full Boltzmann computation to about $0.01\%$, corresponding to $\sim 0.01$~km/s/Mpc in $H_0$ and therefore negligible for all analyses in this paper.

\hicosmo\ also treats neutrinos as massless, with the radiation density fixed by $N_{\rm eff}=3.046$.
For the background distances at $z\lesssim\mathcal{O}(1)$ this is negligible, but in the CMB distance priors it shifts $D_M(z_*)$ relative to the standard Planck analysis with $\sum m_\nu = 0.06$~eV.
The CAMB calibration of section~\ref{sec:architecture} is fitted with the same massless-neutrino background, and the Cobaya runs of section~\ref{sec:cobaya_validation} adopt it as well, so the cross-validation compares the two inference pipelines under identical physics.
Incorporating massive neutrinos into the CMB distance priors is left to future work.

\subsection{Extension to the perturbation level}

The validation in this paper shows that the performance gain requires both the likelihood implementation and the sampling strategy to work together.
Accelerating the likelihood alone is not enough. JIT compilation brings each call to sub-millisecond cost, but a random-walk sampler wastes most of that speed by producing highly correlated samples.
Using NUTS alone is not enough either. NUTS relies on gradients, and if those gradients can only be obtained by finite differences the cost scales linearly with the number of parameters.
\hicosmo\ solves both problems at once, the likelihoods are compilable and differentiable, and NUTS directly consumes the gradients provided by automatic differentiation.

The same conditions apply to the unsolved problem of perturbation-level inference.
The requirement that the architecture of section~\ref{sec:architecture} imposes on a perturbation-level extension is explicit, the forward model must return $C_\ell$, $P(k)$, or compressed observables thereof that are differentiable with respect to the cosmological parameters, after which the likelihood-layer and inference-layer interfaces used in this paper carry over unchanged.
Neural-network emulators such as CosmoPower-JAX \citep{CosmoPower,Piras2024} already provide differentiable approximations of Boltzmann-solver outputs that satisfy this requirement, and the Stage~IV $3\times 2$pt analysis of Piras et al. \cite{Piras2024} has demonstrated the viability of this pattern at the perturbation level on a 157-parameter problem.
The contribution of this paper is to show that the same design operates end to end at the background level, with every step checked directly against mature reference codes and published constraints.
Extending it to the perturbation level requires replacing the model layer with a differentiable Boltzmann surrogate, not redesigning the inference layer.

In summary, the scope of this paper is restricted to background cosmology.
We focus on the background level because its public data, mature reference codes, and published constraints allow every step of the differentiable design to be validated.
Our results show that with an end-to-end differentiable forward model, gradient-based sampling and AD-based Fisher forecasting can operate in a single workflow while remaining consistent with reference analyses.

At present, the \jax\ ecosystem does not yet offer a comprehensive cosmological inference framework comparable to Cobaya in scope. \hicosmo\ is designed to fill this role.
The background-level infrastructure presented here establishes the foundation.
We are currently adding likelihoods for further observational probes, including the TDCOSMO hierarchical analysis of strong-lensing time delays \cite{Birrer2020} and gravitational-wave standard sirens \cite{Jin:2022qnj,Jin:2023sfc,Song:2025bio}.
As differentiable Boltzmann emulators \cite{CosmoPower,Piras2024} and \jax-native cosmological tools \cite{JAXCosmo} continue to mature and the practical demands of Stage~IV surveys grow, we plan to incorporate full perturbation-level likelihoods, extending the framework to weak lensing, galaxy clustering, and CMB power spectrum analyses for next-generation surveys such as DESI, Euclid, LSST, and SKA.
The source code of \hicosmo\ is publicly available\footnote{\url{https://github.com/JingZhaoQi/hicosmo-jax}}, with online documentation at \url{https://jingzhaoqi.github.io/hicosmo-jax}.

\acknowledgments

We acknowledge the developers of the open-source projects \jax, \numpyro, and GetDist \cite{GetDist}.
This work was supported by the National SKA Program of China (Grants Nos. 2022SKA0110200, 2022SKA0110203), the National Natural Science Foundation of China (Grants Nos. 12533001, 12575049, 12473001), the China Manned Space Program (Grant No. CMS-CSST-2025-A02), and the National 111 Project (Grant No. B16009).

\bibliographystyle{JHEP}
\bibliography{main_en}

\end{document}